\documentclass[%
 reprint,
superscriptaddress, nofootinbib,
 amsmath,amssymb,
 aps,
]{revtex4-2}

\usepackage{graphicx}
\usepackage{dcolumn}
\usepackage{bm}
\usepackage{xcolor} 
\usepackage{hyperref} 
\hypersetup{
    colorlinks=true,                
    breaklinks=true,                
    urlcolor= black,                
    linkcolor= blue,                
    bookmarksopen=false,
    filecolor=black,
    citecolor=blue,
    linkbordercolor=red
}
\usepackage[T1]{fontenc}  

\AtBeginDocument{\hypersetup{pdfborder={0 0 1}}}
\newcommand{\specialcell}[2][c]{%
  \begin{tabular}[#1]{@{}c@{}}#2\end{tabular}}

\DeclareGraphicsExtensions{.pdf,.png}

\begin{document}

\preprint{APS/123-QED}

\title{\textbf{Predicting Binary Neutron Star Postmerger Spectra Using Artificial Neural Networks}}

\author{Dimitrios Pesios}
\email{dipesios@auth.gr}
\affiliation{Department of Physics, Aristotle University of Thessaloniki, GR-54124 Thessaloniki, Greece}

\author{Ioannis Koutalios}
\email{koutalios@mail.strw.leidenuniv.nl}
\email{iokoutal@physics.auth.gr}
\affiliation{Department of Physics, Aristotle University of Thessaloniki, GR-54124 Thessaloniki, Greece}
\affiliation{Leiden Observatory, Leiden University, P.O. Box 9513, 2300 RA Leiden, The Netherlands}%


\author{Dimitris Kugiumtzis}
\email{dkugiu@auth.gr}
\affiliation{Department of Electrical and Computer Engineering, Aristotle University of Thessaloniki, GR-54124 Thessaloniki, Greece}%

\author{Nikolaos Stergioulas}
\email{niksterg@auth.gr}
\affiliation{Department of Physics, Aristotle University of Thessaloniki, GR-54124 Thessaloniki, Greece}%


\date{\today}

\begin{abstract}

Gravitational waves in the postmerger phase of binary neutron star mergers may become detectable with planned upgrades of existing gravitational-wave detectors or with more sensitive next-generation detectors.  The construction of template banks for the postmerger phase can facilitate signal detection and parameter estimation. Here, we investigate the performance of an artificial neural network in predicting simulation-based waveforms in the frequency domain (restricted to the magnitude of the frequency spectrum and to equal-mass models) that depend on three parameters that can be inferred through observations, neutron star mass, tidal deformability, and the gradient of radius versus mass. Compared to a baseline study using multiple linear regression, we find that the artificial neural network can predict waveforms with higher accuracy and more consistent performance in a cross-validation study. We also demonstrate, through a recalibration procedure, that future reduction of uncertainties in empirical relations that are used in our hierarchical scheme will result in more accurate predicted postmerger spectra.

\end{abstract}

\maketitle


\section{INTRODUCTION}
\label{seq:Intro}

During the previous decade, we witnessed the birth of a new branch of observational astronomy that does not rely on the electromagnetic spectrum to derive observations of the universe and its components \cite{Abbott_2016blz}. Instead, it uses the properties of space-time itself to reveal hidden aspects of astronomical objects by means of gravitational wave (GW) interferometric detectors. These space-time properties are exerted by violent events occurring in the universe, such as the coalescence of binary black hole (BBH) or binary neutron star (BNS) systems, leading to the generation of detectable gravitational waves. 
In the first three observing runs, 90 events have been reported by the LIGO-Virgo-KAGRA (LVK) collaboration, the majority of them concerning BBH systems and only two of them concerning BNS systems \cite{LIGOScientific_2021djp}. More GW candidate events are accumulating in the ongoing O4 phase that started in May 2023. As current detectors are being improved and third-generation detectors are being designed, a greater number of signals, and possibly even new types, are expected to be detected in the near future. \cite{KAGRA:2020npa,2021NatRP...3..344B,https://doi.org/10.48550/arxiv.2104.02445}. 

One of the anticipated discoveries, which so far has remained undetected \cite{Abbott:2017dke, grace2024gravitational}, is the gravitational wave signal from the postmerger phase of BNS mergers. However, these signals may be identified using planned upgrades that extend beyond the fifth observing run (O5) and become more probable with third-generation or dedicated high-frequency detectors \citep{PhysRevD.90.062004, Chatziioannou_2017,Torres_Rivas_2019,BNS_3G_1,BNS_3G_2,BNS_detection2,BNS_detections,CE_BNS_2022, 2022arXiv220509979B,HF}. 

The post-merger phase may last up to a few tens of milliseconds and is characterized by the emission of significant amounts of gravitational radiation at distinct frequencies of a few kilohertz. Identifying specific post-merger frequencies could set strong constraints on the radius of neutron stars and, consequently, on their equation of state (EOS), which is a major goal in high-energy astrophysics, see \cite{2016EPJA...52...56B,Baiotti_2017,Bauswein_2019,Baiotti2019, Friedman_2020,Bernuzzi2020,Diedrich2021mar,Sarin2021jun,PhysRevLett.128.161102} and references therein.

\begin{figure}[ht!]
    \centering
    \includegraphics[width=\linewidth]{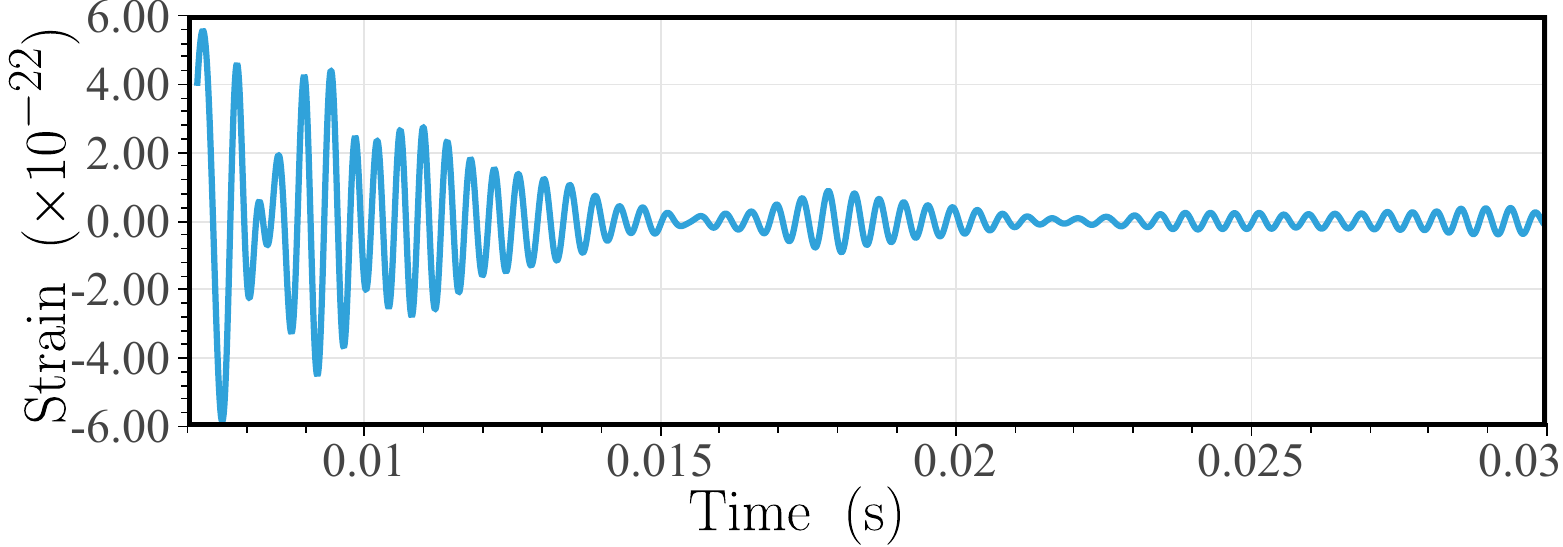}
    \centering
    \includegraphics[width=\linewidth]{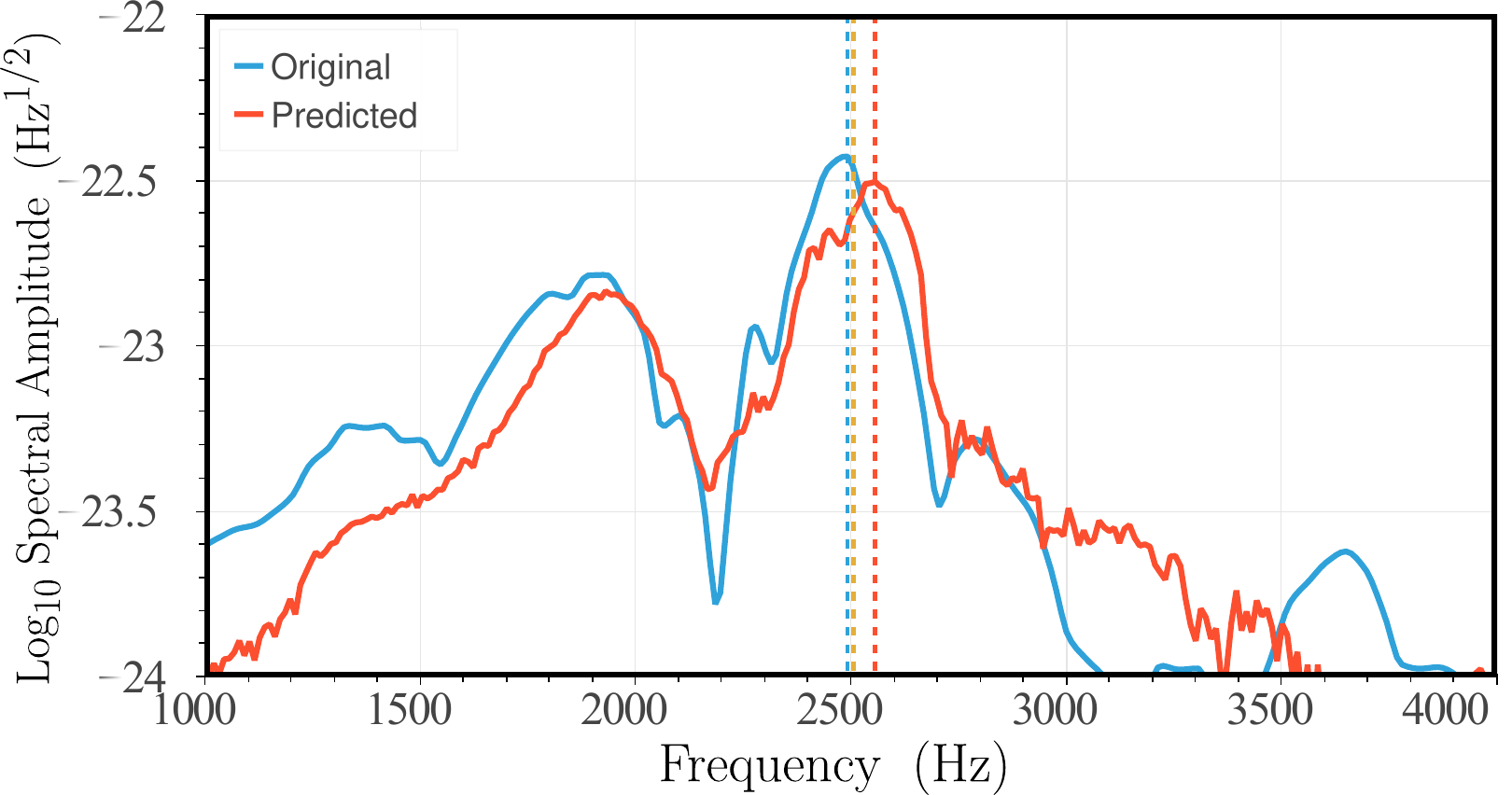}
    \caption{Plus polarization of the postmerger strain, $h_{+}$, in the time domain (top panel) and the frequency domain (bottom panel) for an 1.3$M_\odot$ equal-mass BSN merger with the DD2 EOS. The source is assumed to be at 50 Mpc. The spectrum with our ANN-based method (red curve) agrees well with the original spectrum (blue curve). The blue dashed vertical line and the red one indicate the frequencies where each spectrum attains its peak value correspondingly, whereas the orange one indicates the predicted frequency of the empirical relation.}
    \label{fig:DD2}
\end{figure}

Figure \ref{fig:DD2} shows an example of the GW strain during the postmerger phase and the corresponding Fourier spectral amplitude for a particular EOS (DD2) and component mass 1.3$M_\odot$ (the same for both stars), extracted from the CoRe v2 dataset \cite{Gonzalez_2022mgo}. 
During merger, linear nonaxisymmetric oscillation modes, nonlinear combination tones, and other transient effects are excited in the remnant; see e.g. \citep{Stergioulas_2011,Hotokezaka_2013,Bauswein_2015,Takami_2015,Clark_2016,PhysRevD.94.064011,De_Pietri_2018,Fields:2023bhs}. The main oscillation mode is the $l=m=2$ $f$-mode, commonly\footnote{In some cases it is also denoted as $f_2$.} referred to as $f_{\rm{peak}}$. For comprehensive reviews of the various characteristics of the post-merger spectrem, see 
\cite{2016EPJA...52...56B,2019AIPC.2127b0013B,2019JPhG...46k3002B,Sarin2021jun,Friedman_2020,annurev-nucl-013120-114541}.

The post-merger GW spectrum (typically in the range between $\mathrm{1.5~kHz}$ and $\mathrm{4~kHz}$) depends on the masses of the two binary components and the EOS, see, e.g. ~\citep{PhysRevLett.94.201101,PhysRevD.71.084021,PhysRevLett.99.121102,Bauswein_Janka, Bauswein_2012,Hotokezaka_2013,Takami_2014,Takami_2015,Friedman_Stergioulas_2020,Fields:2023bhs} and includes information that may allow the inference of various neutron star properties, see, e.g. ~\cite{1992ApJ...401..226R,PhysRevLett.94.201101, PhysRevD.90.062004,PhysRevD.93.124051, PhysRevD.93.044019,Haster_2020,2023arXiv231010728T}. This could be achieved, for instance, using empirical relations that relate $f_{\rm peak}$ to radii of non-rotating neutron stars in the inspiral phase~\cite{Bauswein_2012, Hotokezaka_2013, Takami_2015, Bernuzzi_2015, Bose_2018, Bauswein_2019, PhysRevD.100.104029, Vretinaris_2020,Criswell_2023}. 

A prerequisite for extracting source parameters from future detections of the postmerger signal is the availability of a specialized template bank that depends on the two component masses and the EOS. However, BNS simulations in numerical relativity are time-consuming and, at the moment, only a few hundred different simulations have been performed. Even in the case of BBH mergers, where a larger number of numerical simulations are available, regression methods have been employed to construct template banks to reach the number of templates required for accurate detection and parameter estimation; see \cite{FrankOhme2020} and references therein. In the case of BNS mergers, the problem is more challenging. Ideally, we would like to construct detailed template banks that describe and parametrize the full inspiral-merger-postmerger waveform and encode the impact of the EOS.

In \cite{krolak2023search} a mathed-filtering approach for detecting the postmerger signal is described. The template bank consists of the analytic waveform model proposed in \cite{Bose_2018}. Additionally, in \cite{stad1556} the impact of transient noise artifacts with high-frequency components on the sensitivity of the search for GWs in the postmerger phase was considered.  In \cite{breschi2022numerical}  a frequency-domain analytical model was explored using wavelets in combination with empirical relations, where also a recalibration was applied and its feasibility for detecting the postmerger phase with third-generation detectors was studied in \cite{breschi2022kilohertz}. In addition, \cite{Puecher_2023} presented another analytic model in the frequency domain, based on a parameterized Lorentzian function,  for the complete coalescence process. Apart from individual detections, the possibility of combining information from a number of weak postmerger signals (too weak to be detectable individually) has been explored in  \cite{Criswell_2023}, concluding that first indications for the presence of postmerger GW emission could come with the final updgrades of the second-generation detectors. Lastly, in \cite{Tringali_2023} a morphology-independent method was presented to characterise the fate of the post-merger remnant..

A large number of detections of the inspiral phase of BNS mergers using third-generation detectors is expected to lead to tight EOS constraints; see \cite{iacovelli2023nuclear} and reference therein. However, using only the inspiral phase, one cannot probe EOS properties at densities higher than those encountered in the core of isolated neutron stars before they merge. In contrast, the detection of GWs in the postmerger phase, will allow us to probe higher densities and test for the presence of phase transitions or exotic components \cite{2020PhRvD.102l3023B,PhysRevLett.122.061101,PhysRevLett.122.061102,Most2020,PhysRevD.104.083029,breschi2023prepostmerger,2024PhRvD.109d3015B} or  deviations from general relativity \cite{PhysRevD.87.081506,PhysRevD.89.044024,PhysRevD.89.084005,PhysRevD.97.064016,PhysRevD.104.104036,PhysRevD.106.104055,PhysRevD.108.064057,PhysRevLett.128.091103,PhysRevD.108.024058,PhysRevD.108.063033}.

Easter et al.  \cite{PhysRevD.100.043005} introduced a methodology comprising a hierarchical model of two main steps to predict postmerger spectra in the frequency domain. Here, we construct a similar model, but, in contrast to \cite{PhysRevD.100.043005}, we do not use an empirical relation for the compactness $C=M/R$, but, instead, we use the inverse gradient $dR/dM$ of the mass-radius relation (where $M$ is the gravitational mass and $R$ the circumferential radius of a relativistic star).
Furthermore, we extended the training set from 35 spectra in \cite{PhysRevD.100.043005} to a total of 87 spectra. Apart from improved results with this hierarchical multivariate regression scheme, we are also interested in exploring the
application of artificial neural networks (ANNs) as a regression scheme for predicting postmerger spectra. To this end, we train a feed-forward ANN on the same data and perform extensive comparisons between the two methods. Although, at the moment, the training set comprises a limited number of models, we obtain comparable or slightly better results with the ANN regression, showing that this is a very promising method to be applied in future searches, when many more simulation results will be available. Furthermore, using a recalibration of the results, we show that if the uncertainties of the employed empirical relations will be reduced in the future (due to better EOS constraints obtain through GW observations in the inspiral phase) this will greatly improve the accuracy of our proposed method.

This work is organized as follows: In Section \ref{section_ii}, we present the two regression-based techniques used to construct predictive models, along with some statistical and ML-based concepts. In Section \ref{section_iii} 
we outline the experiments conducted to assess the performance of the models, such as comparing the resulting Fitting Factor histograms before and after recalibration and a necessary cross-validation sensitivity analysis. Finally, in Section \ref{section_iv} we make some final remarks and draw conclusions. Some complementary results are depicted in collective spectra plots in Appendix \ref{AppendixA} for the two models.

We note that in certain expressions, such as $M/R$ we use the standard convention of setting $c=G=1$.

\section{PREREQUISITES} \label{section_ii}

Here, we introduce the applied machine learning and statistical techniques used to construct and analyze the predictive models of our work. Waveform model-based prediction is useful to construct templates that cover a wide range of possible system parameters. The statistical aspect of predictive modeling is primarily concerned with minimizing the error of a model in order to make predictions of future outcomes as accurately as possible.

\subsection{Regression techniques for prediction}

We mainly leverage two approaches; a classical multi-input and multi-target linear regression approach to formulate a multivariate linear model and, secondly, a more general function approximating approach, based on an Artificial Neural Network model, with various enhancements. Both approaches can fall into the regression-based description, since they obey the same mathematical equation (as in Eq. \ref{1} below) connecting input and output, however, it is implemented differently in the two approaches.

\subsubsection{Multivariate linear regression (MLR)}

Multivariate linear regression \textcolor{black}{is an extension of multiple linear regression allowing for more than one output variables, which may be correlated} \cite{Izenman2008, foxweisberg}. The multivariate linear model is:
\begin{equation} \label{1}
    \underset{(n \times m)}{Y} = \underset{(n \times l+1)}{X} \underset{(l+1 \times m)}{B} + \underset{(n \times m)}{E} \; ,
\end{equation}
\noindent
where $Y$ is a matrix of $n$ observations on $m$ response variables (regressands), $X$ is a model (design) matrix with columns for $l + 1$ regressors, typically including a column of ones for the regression constant, $B$ is a matrix of regression coefficients, with one column for each response variable, and $E$ is a matrix of residuals \textcolor{black}{, where the rows $E_i$, $i=1,\ldots,n$, have all mean 0 and the same error covariance matrix, and they are uncorrelated to each other}.

\textcolor{black}{Assuming that $X^{\prime}X$ is non-singular and the residuals are Gaussian, it can be shown that} the maximum-likelihood estimate of $B$ in the multivariate linear model is equivalent to \textcolor{black}{the least squares estimate} for the individual responses:
\begin{equation} \label{2}
    \hat{B} = (X'X)^{-1}X'Y \;\; ,
\end{equation}
\noindent
where the hat indicates the estimated regression values of the coefficients. \textcolor{black}{If the responses are assumed to be uncorrelated, the} multivariate linear regression can be \textcolor{black}{treated} as many independent and stacked multiple linear regressions.

\subsubsection{Artificial Neural Networks as universal approximators}

Artificial neural networks, in contrast, are more general problem-solving \textcolor{black}{models}. \textcolor{black}{For the multivariate regression problem, ANNs estimate any function $f: R^{k} \rightarrow R^{m}$, having as input a vector of $k$ regressors and as output a vector of $m$ responses}. \textcolor{black}{The formulation of a regression solution using an ANN bears a mathematical expression, which is complicated, with the complication increasing with the number of layers. In this expression, crucial non-linearity is introduced by means of specialized activation functions applied to the outputs of individual neurons or layers within the network.}

Although an ANN may consist of more than three layers of neurons, it has been mathematically shown that any functional relationship can, in practice, be implemented with a three-layer ANN \cite{goodfellow2016deep, HORNIK1989359, HORNIK1991251}. Theoretically, according to Universal Approximation Theorem, a network with just one hidden layer and non-linear activation functions should be sufficient to model any function mapping, but in practice, additional layers are included. In our implementation of ANN,  we use four layers and make further enhancements to improve network performance.

The first enhancement we used was a \textit{learning rate} (LR) \textcolor{black}{scheduling} scheme \cite{https://doi.org/10.48550/arxiv.1212.5701,https://doi.org/10.48550/arxiv.1412.7419,https://doi.org/10.48550/arxiv.1908.03265}, involving a non-manual tuning of the LR value. Typically, we want the LR value not to be too low, since the neural network would learn slowly, but also not too high, since this would result in divergence issues. \textcolor{black}{The strategy we followed suggests starting with} a low LR (warm-up), which gradually increases (LR annealing), followed by a slow LR decrease for a certain number of steps when the loss function reaches a plateau, to achieve the best possible loss value (local minima). Moreover, introducing a warm-up period in the LR can reduce the initial variance of the training procedure and further stabilise it.

Another enhancement, which is common in statistics, is \textit{standardization},  (\textcolor{black}{that is, subtracting the mean and dividing with the standard deviation}) of the input values fed into the neural network \textcolor{black}{(and the same for the output values during training)}, and the reverse procedure of bringing the \textcolor{black}{predicted} output values back into \textcolor{black}{the original domain}. For \textcolor{black}{each variable, the mean and standard deviation are computed} from the input/output training samples and divide by the respective standard deviation. The same mean and standard deviation are then used in \textcolor{black}{the reverse procedure on the output values.}.

Lastly, one can use \textit{early stopping} while training the ANN, a technique that stops the training procedure when a low plateau is reached in the corresponding validation loss, calculated using a separate set other than the training one. In this way, the prediction accuracy and the training time can improve. Furthermore, to ensure that the ANN is able to generalise and mitigate overfitting effects in deployment, other techniques can be employed, such as noise addition in the output of constituting layers \cite{6796505} and random dropout of nodes \cite{SIETSMA199167,JMLR:v15:srivastava14a}.

\begin{table*}
\caption{
Description of the training data set of 87 waveforms. The first column lists the EOS name, the second column lists the number of different models for each EOS, the third column gives the main reference for each case and the last column list the component masses of the individual equal-mass models for each EOS. For the APR4, H4 and SLy, the Core v2 database was used above 1.325 $M_\odot$.}
\label{tab:table1}
\begin{ruledtabular}
\resizebox{\textwidth}{0.15\textheight}{%
\begin{tabular}{cccc}
{\bf EOS}  & {\bf Waveforms} & {\bf References} & {\bf Component Masses} (in $M_\odot$)\\
\hline
ALF2 & 10 &Rezzolla \& Takami \cite{Takami_PhysRevD.93.124051} and CoRe v2 \cite{Gonzalez_2022mgo}& \specialcell{1.2, 1.225, 1.25, 1.275, 1.3, 1.325,\\ 1.35, 1.3505, 1.375, 1.3755}\\ \hline
APR4 & 7 & Rezzolla \& Takami \cite{Takami_PhysRevD.93.124051} & 1.2, 1.225, 1.25, 1.275, 1.3, 1.325, 1.35 \\ \hline
BHBlp & 4 & CoRe v2 \cite{Gonzalez_2022mgo} & 1.25, 1.3, 1.35, 1.4 \\ \hline
BLh & 4 & CoRe v2 \cite{Gonzalez_2022mgo} & 1.3, 1.3325, 1.364, 1.4 \\ \hline
DD2 & 7 & CoRe v2 \cite{Gonzalez_2022mgo} & 1.2, 1.25, 1.3, 1.35, 1.364, 1.4, 1.5 \\ \hline
ENG & 1 & CoRe v2 \cite{Gonzalez_2022mgo} & 1.3495 \\ \hline
GNH3 & 7 & Rezzolla \& Takami \cite{Takami_PhysRevD.93.124051} & 1.2, 1.225, 1.25, 1.275, 1.3, 1.325, 1.35 \\ \hline
H4 & 13 &Rezzolla \& Takami  \cite{Takami_PhysRevD.93.124051} and CoRe v2 \cite{Gonzalez_2022mgo}& \specialcell{1.2, 1.225, 1.25, 1.275, 1.3, 1.325, 1.3495\\ 1.35, 1.3505, 1.3715, 1.3725, 1.3735, 1.3795}  \\ \hline
LS220 & 4 & CoRe v2 \cite{Gonzalez_2022mgo} & 1.2, 1.35, 1.364, 1.4 \\ \hline
MPA1 & 8 & Soultanis, Bauswein \& Stergioulas \cite{PhysRevD.105.043020} & 1.2, 1.25, 1.3, 1.35, 1.4, 1.45, 1.5, 1.55 \\ \hline
MS1 & 2 & CoRe v2 \cite{Gonzalez_2022mgo} & 1.3495, 1.351\\ \hline
MS1b & 8 & CoRe v2 \cite{Gonzalez_2022mgo} & 1.35, 1.3505, 1.375, 1.3805, 1.381, 1.5, 1.6, 1.7 \\ \hline
SFHo & 2 & CoRe v2 \cite{Gonzalez_2022mgo} & 1.35, 1.364 \\ \hline
SLy & 10 &Rezzolla \& Takami \cite{Takami_PhysRevD.93.124051} and CoRe v2 \cite{Gonzalez_2022mgo}& \specialcell{1.2, 1.225, 1.25, 1.275, 1.3, 1.325,\\ 1.35, 1.351, 1.3575, 1.364} \\ 
\end{tabular}%
}
\end{ruledtabular}
\end{table*} 

\subsection{Statistical design of experiments}

In this subsection, we \textcolor{black}{briefly} describe some statistical concepts used in our work to efficiently design statistical experiments, such as hierarchical or multilevel regression models and unbalanced designs. Reference \cite{montgomery2008design} provides a thorough and general discussion on the subject.

\subsubsection{Hierarchical or multilevel models}

\textcolor{black}{Some data settings require a} a hierarchy of regressions, the so-called hierarchical linear models (also known as multilevel models), for instance, by regressing $X$ on $Y$ and $Y$ on $Z$ (a two-level approach). 

These kinds of designs are called hierarchical for two reasons \cite{GelmanHill:2007}: first, due to the structure of the data, \textcolor{black}{and} second, because of the model itself, which may demonstrate its own hierarchy with within-group regression parameters controlled by hyperparameters of some upper-level model. Depending on whether these parameters are considered as random variables or not, we distinguish between random-effects and fixed-effects models.

\subsubsection{Balanced vs. unbalanced design}

During the design of a regression solution, data usually form groups based on a common characteristic, as frequency series belonging to the same equation of state (EOS) as far as this work is concerned.

When there is an equal number of observations in each group, we can infer that the design matrix of the regression will be homogeneous and its values (in our case columns) evenly distributed, namely \textit{balanced}. Conversely, when there is not an equal number of observations in each group, the design will be inhomogeneous, something that will lead to an \textit{unbalanced} design matrix.

Balanced designs are often preferred over unbalanced ones, due to their higher statistical power and reliable test statistic. However, in many practical problems this is not a strong constraint on the applicability of a solution, such as linear regression.

\begin{figure*}[t!]
\centering
\begin{minipage}{0.33\textwidth}
  \centering
  \includegraphics[width=\linewidth]{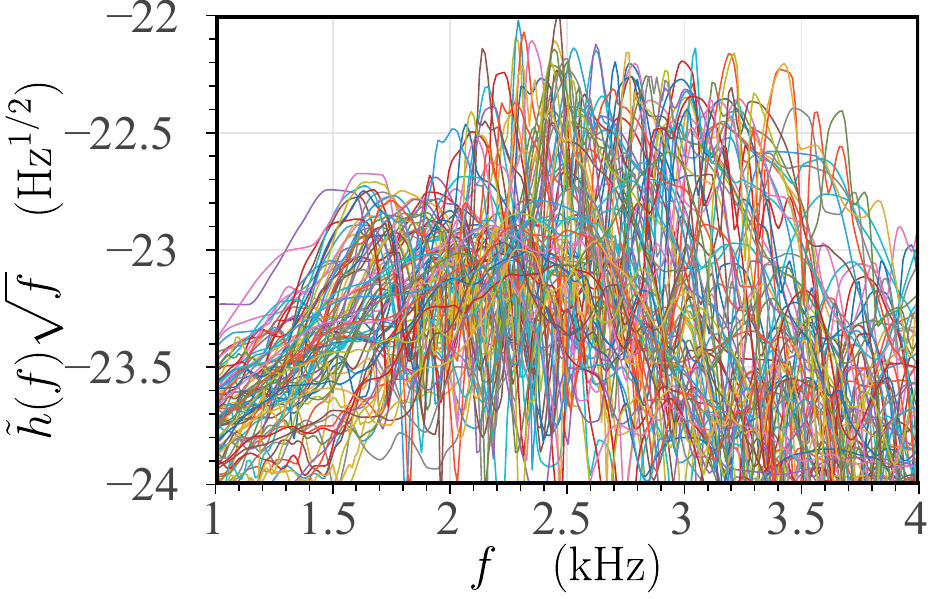}
\end{minipage}%
\begin{minipage}{0.33\textwidth}
  \centering
  \includegraphics[width=\linewidth]{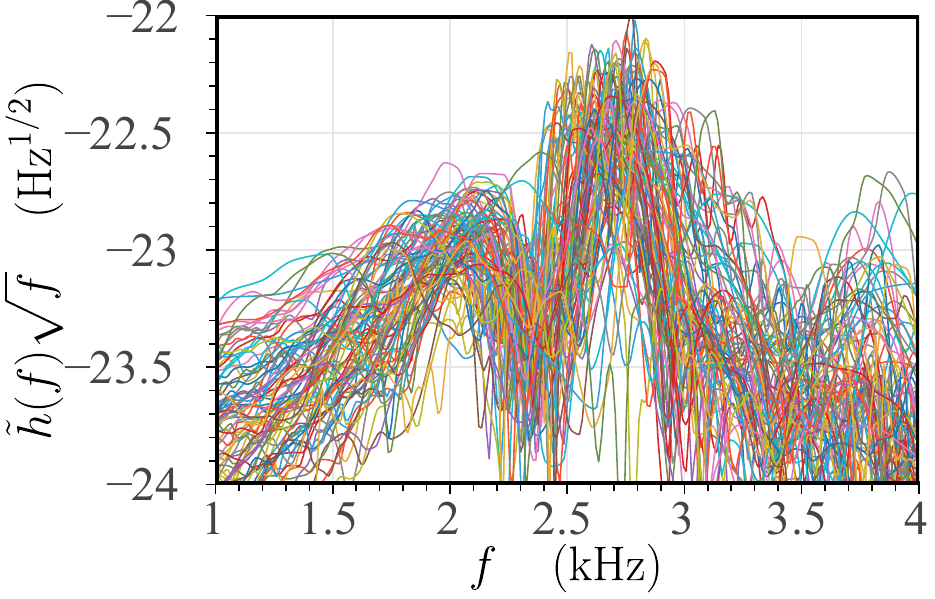}
\end{minipage}
\begin{minipage}{0.33\textwidth}
  \centering
  \includegraphics[width=\linewidth]{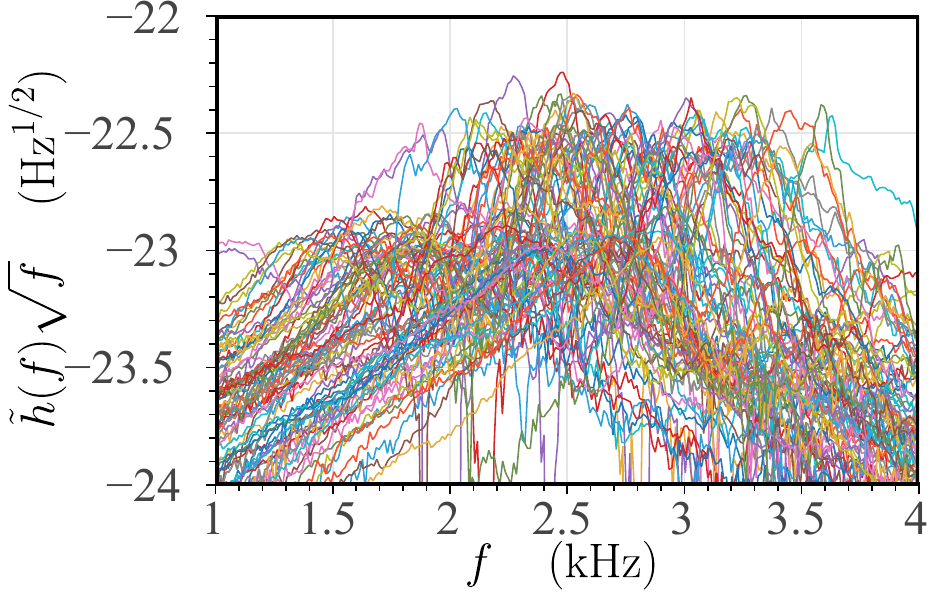}
\end{minipage}
\caption{{\it Left panel}: 
 Collective plot of the 87 different postmerger spectra used in both the MLR regression and the ANN method. {\it Middle panel}: The same spectra, but partially aligned to a common $f_{\rm peak}$ frequency, using the empirical relation of Eq. (\ref{eq:empiricalVret}) (departures from a perfect alignment are due to uncertainties in the empirical relation). {\it Right panel}: Corresponding predicted spectra, using the ANN-based method, after recalibration (see the text for details).}
\label{fig:spectra_two}
\end{figure*}

\subsection{The fitting factor}

To compare \textcolor{black}{the predicted waveform, denoted $h_1(t)$,} with the \textcolor{black}{corresponding spectra of the original waveform $h_2(t)$ using our methodology}, we use a noise-weighted index called the fitting factor (FF), which denotes their overlap, defined as:
\begin{equation}
    F F\left(h_{1}, h_{2}\right) \equiv \frac{\left\langle h_{1} \mid h_{2}\right\rangle}{\sqrt{\left\langle h_{1} \mid h_{1}\right\rangle\left\langle h_{2} \mid h_{2}\right\rangle}} \; ,
\end{equation}
\noindent
where the (approximate) inner product is (as introduced in \cite{PhysRevD.100.043005}):
\begin{equation}
    \left\langle h_{1} \mid h_{2}\right\rangle \equiv 4 \int_0^\infty d f \frac{\left|\tilde{h}_{1}(f)\right|\left|\tilde{h}_{2}(f)\right|}{S_{h}(f)},
\label{eq:inner}
\end{equation}
and $\tilde{h}(f)$ is the Fourier transform of a waveform $h(t)$.
In Eq. \ref{eq:inner}, $S_{h}(f)$ is the noise power spectral density (PSD), averaged over a sufficiently long time segment. We will assume that the PSD is a stationary, colored Gaussian noise model. A good match (or overlap) between the two spectra is indicated with values \textcolor{black}{of FF} close to 1.

We note that this definition of FF was introduced in \cite{PhysRevD.100.043005}, since we are only dealing with the amplitudes (and not the phases) of the \textcolor{black}{Fourier transform} and differs from the full definition introduced in \cite{PhysRevD.52.605}.

\subsection{Cross-validation}

Cross-validation (CV), or out-of-sample testing, is one of the most popular approaches for model evaluation \cite{hastie01statisticallearning}. Usually, \textcolor{black}{the $k$-fold CV is partitioning} the data set into $k$ roughly equal parts, training the model with $k-1$ of them and testing it using the one excluded. This procedure \textcolor{black}{is repeated} $k$ times, \textcolor{black}{covering, in this way, the entire dataset}.

A subcase of $k$-fold CV is the so-called {\it leave-one-out} CV (LOO-CV), when $k$ equals the cardinality of the dataset. This approach is employed in \cite{PhysRevD.100.043005}.
However, various trade-offs emerge using a $k$-fold CV and many of them are still under debate \cite{journals/jmlr/BengioG04,ZHANG201595,10.5555/1643031.1643047} with one of the most famous being the bias-variance trade-off \textcolor{black}{with respect to $k$ value}. Thus, in this work, we will conduct a detailed $k$-fold cross-validation study, \textcolor{black}{varying $k$ from 2 to 87}.

\section{EXPERIMENTS} \label{section_iii}

In this section, we describe the experiments conducted to \textcolor{black}{assess} the robustness and efficiency of our proposed extended methodology. \textcolor{black}{First, the dataset is described. Then,  implementation-specific details about the MLR and ANN regression models are given, and finally the two models are compared.}

\subsection{Dataset description}

For both regression-based models, we have to provide data both for the regressors (input) and the regressands (output). The columns of the design matrix are populated with data pertaining to three regressors ($k=3$ in Eq. \ref{1}), which are the component mass $M$ in an equal-mass ($q=1$) binary system,  the tidal coupling constant $\kappa_2^\tau$ (expressing tidal deformability), and the derivative $dR/dM$, which corresponds to the inverse gradients of the $M(R)$ relation of nonrotating equilibrium models for a given EOS. In the case of the MLR model, the design matrix includes a first row consisting of unit values. \textcolor{black}{The response matrix $Y$ is formed by the amplitudes of the postmerger waveforms in the Fourier domain.}


In Table \ref{tab:table1} we list a detailed description of the 87 different waveforms used in our study. Most of the waveforms are from the  CoRe v2 database \cite{Gonzalez_2022mgo} (47 waveforms) and the Rezzola and Takami catalogue \cite{Takami_PhysRevD.93.124051} (32 waveforms), while 8 waveforms are from Soultanis et al. \cite{PhysRevD.105.043020}. All waveforms were scaled to a common distance of 50 Mpc truncated at their maximum amplitude to separate the inspiral and postmerger phases. The postmerger part was then \textcolor{black}{transformed} to the Fourier domain and the resulting spectra were partially aligned using the empirical relation of Eq. \ref{eq:empiricalVret}.

Lastly, we confined the frequency range of the resulting spectra to the astrophysically-relevant domain of 1-4kHz, leaving some  tolerance due to the shift-unshift procedures (depending on the maximum displacement of the spectrum). The total number of frequency bins \textcolor{black}{is $m=370$}.

\begin{figure*}
\centering
\begin{minipage}{0.45\textwidth}
  \centering
  \includegraphics[width=\linewidth]{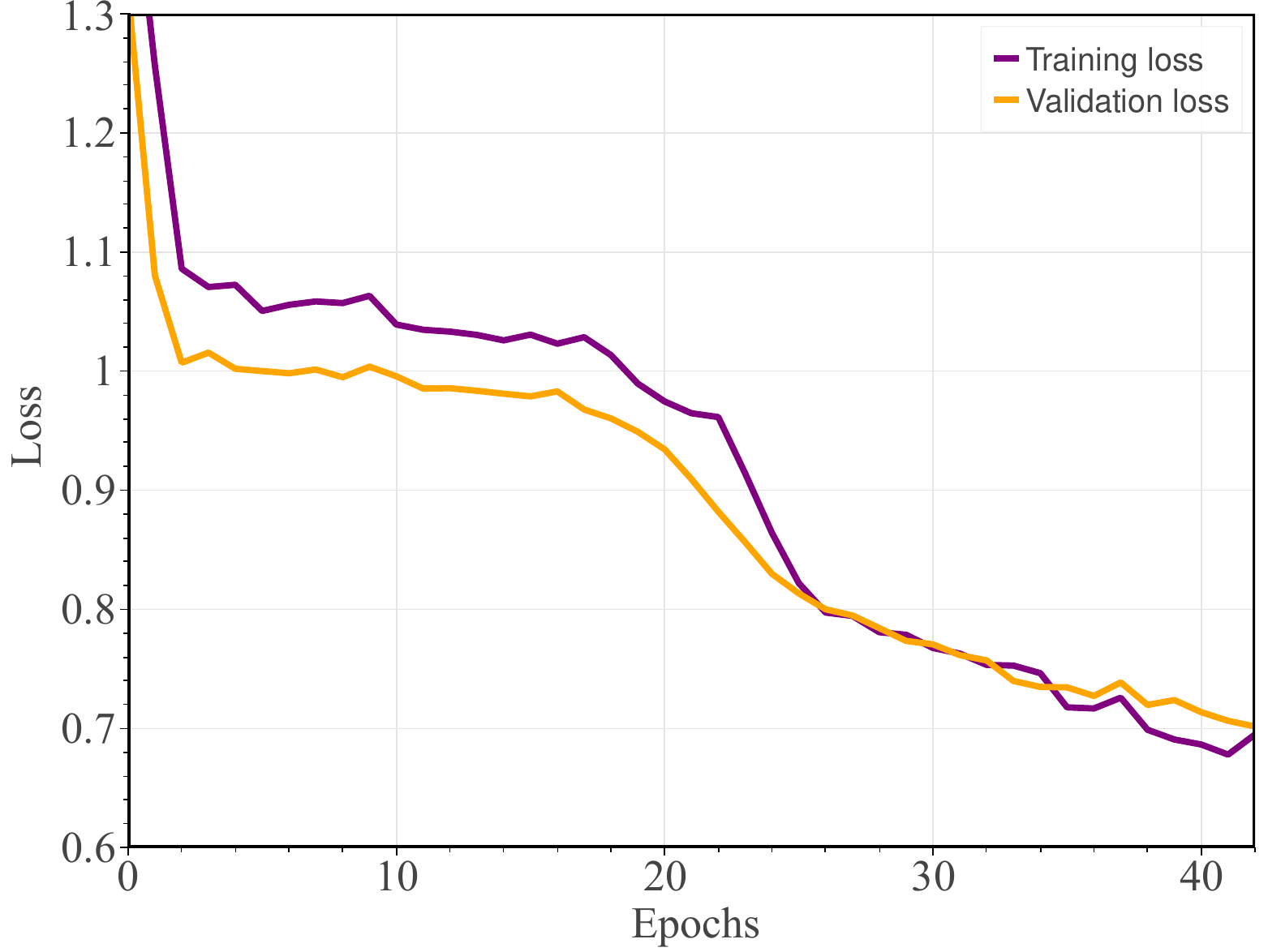}
\end{minipage}%
\begin{minipage}{0.45\textwidth}
  \centering
  \includegraphics[width=\linewidth]{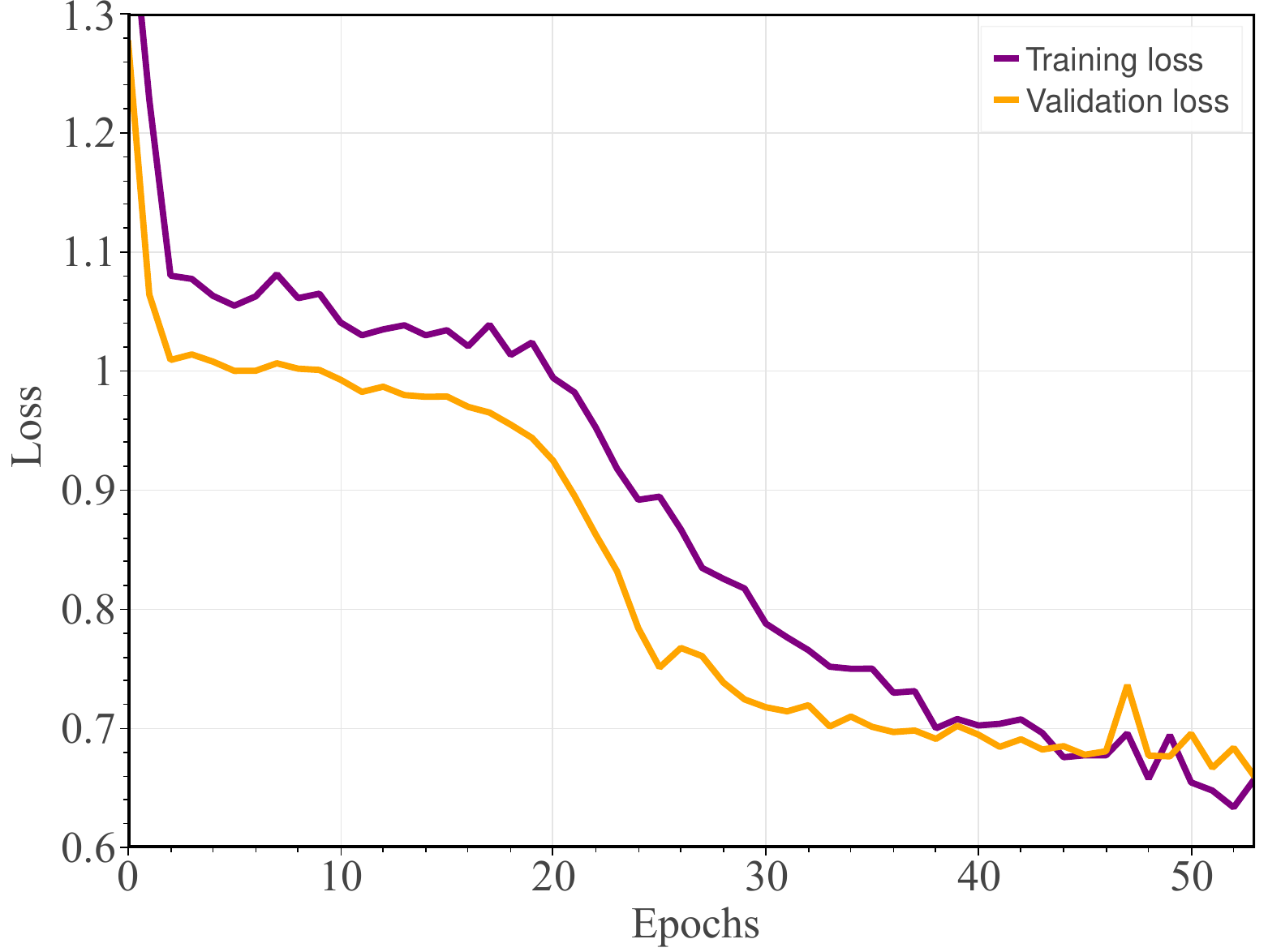}
\end{minipage}
\begin{minipage}{0.45\textwidth}
  \centering
  \includegraphics[width=\linewidth]{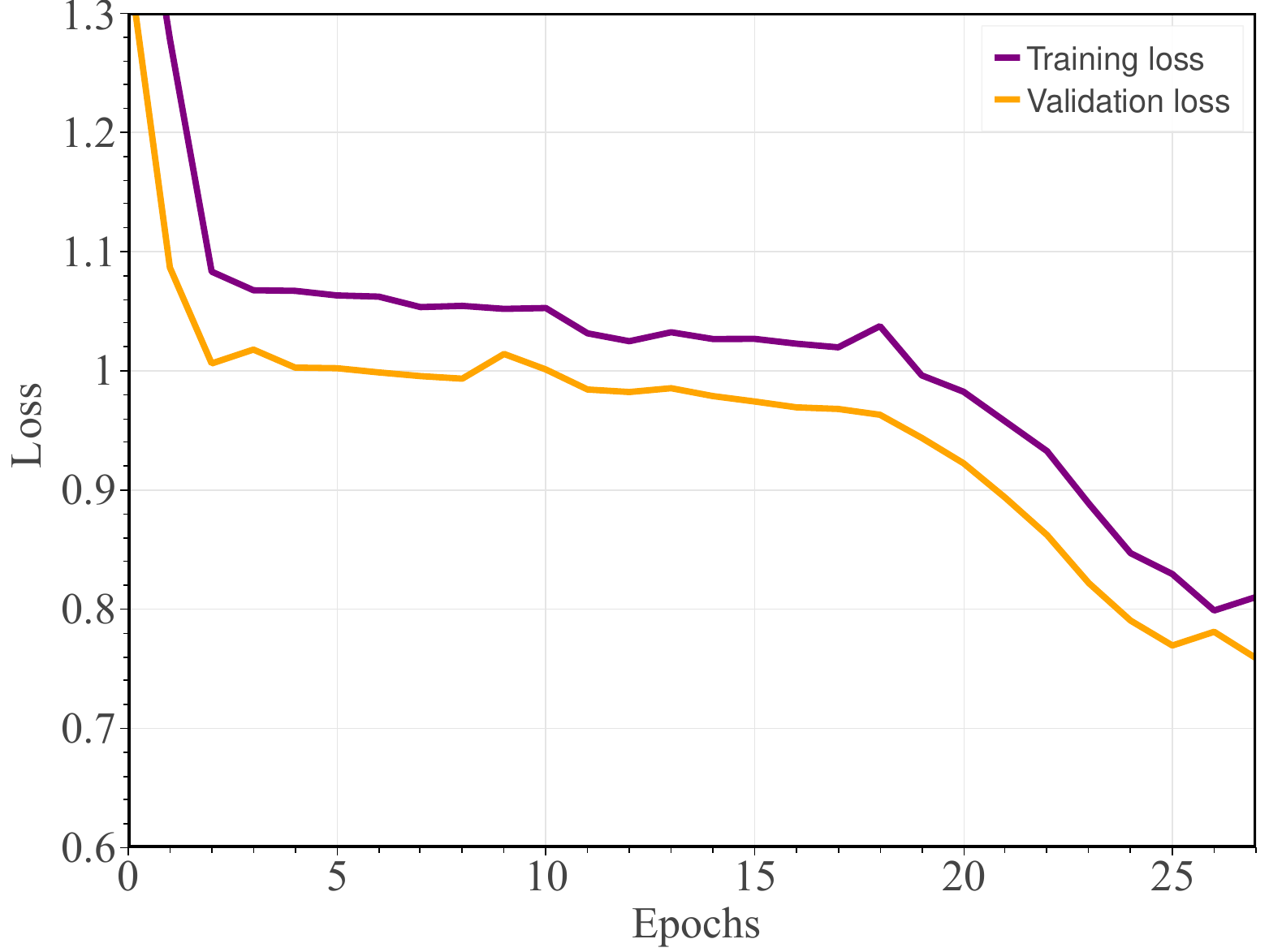}
\end{minipage}%
\begin{minipage}{0.45\textwidth}
  \centering
  \includegraphics[width=\linewidth]{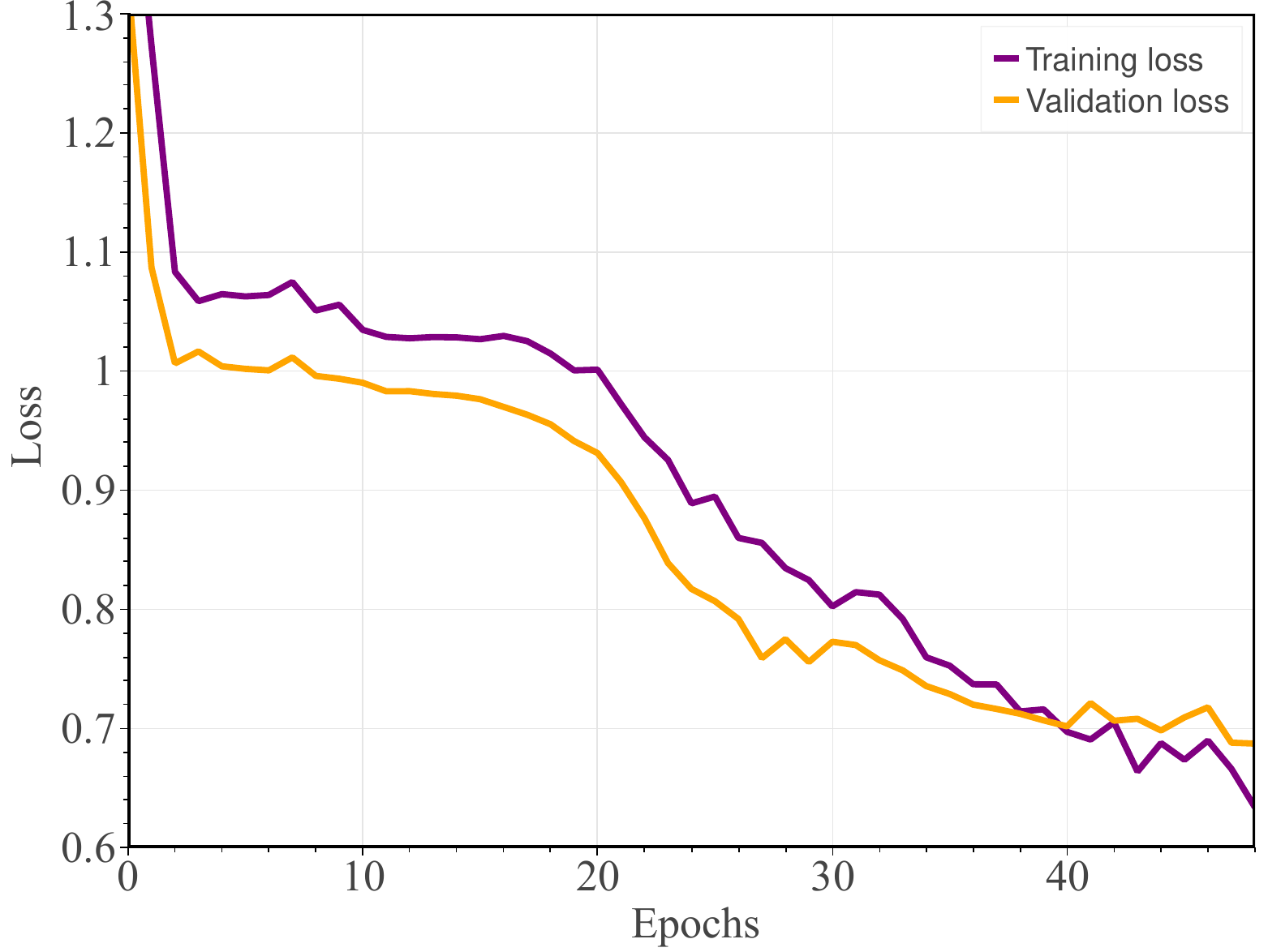}
\end{minipage}%
\caption{Training and validation loss as a function of the number of epochs, shown \textcolor{black}{for four} different shufflings of the training set. The training was stopped when the validation loss rose above the training loss for seven consecutive epochs.}
\label{fig:all_loss_curves}
\end{figure*}

\subsection{Alignment of postmerger spectra}

A common feature of all postmerger spectra (irrespective of the source parameters) is that they are dominated by the $f_{\rm peak}$ frequency. In \cite{Clark_2016}, this feature was used to align the spectra to a common reference frequency, which was useful for applying a principal component analysis (PCA). Similarly, the hierarchical method of  \cite{PhysRevD.100.043005} includes an alignment step, which needs to be reversed after the regression is performed. Here, we also align the spectra to a common reference frequency, which we take to be the average value of $f_{\rm peak}$ in our dataset. \textcolor{black}{For given values of the tidal coupling
constant $\kappa_2^{\tau}$ and component mass $M$ for an equal-mass system, we predict $f_{\mathrm{peak}}$ by the empirical relation }
\begin{equation}
    f_\mathrm{peak}(\kappa_2^\tau, M) = 4\frac{\beta_{1}}{M}\mathrm{ln}\left( \frac{\beta_{0}}{8{\kappa_{2}^\tau}} \right).
    \label{eq:empiricalVret}
\end{equation}
\noindent
Eq. (\ref{eq:empiricalVret}) \textcolor{black}{is based on} the empirical relation Eq. (41) in \cite{PhysRevD.101.084039_Vretinaris}, but is written here \textcolor{black}{differently} and using the known relation between mass-weighted tidal deformability $\tilde \Lambda$ and $\kappa_2^\tau$, as well as the relation between chirp mass $M_{\rm chirp}$ and component mass $M$ for equal mass systems. The fitting values for $\beta_0$ and $\beta_1$ \textcolor{black}{were re-calculated, since a single composite variable was used in \cite{PhysRevD.101.084039_Vretinaris}}.

The left panel in Fig.  \ref{fig:spectra_two} shows a collective plot of the 87 different postmerger spectra we use in our study. The value of $f_{\rm peak}$ can be as low as $\sim 1.5$ kHz, or as high as $\sim 3.5$ kHz. The middle panel in Fig. \ref{fig:spectra_two} shows the same spectra, but aligned using the empirical relation in Eq. (\ref{eq:empiricalVret}). Due to the uncertainties of the empirical relation, the alignment is only partial, i.e. the spectra are not perfectly aligned according to their actual peak values. However, the empirical relation allows the alignment to be reversed after the regression is performed.

The partially aligned spectra, in the middle panel of Fig. \ref{fig:spectra_two} serve as  training data for both the MLR and ANN regression models \textcolor{black}{presented} in Section \ref{sec:models}.  The right panel in Fig.  \ref{fig:spectra_two} shows the corresponding predicted spectra (with the alignment reversed), using the ANN-based method and after applying the recalibration procedure discussed in Section \ref{sec:recalibration}.

\subsection{Regression models}
\label{sec:models}

    In this subsection, we describe in detail the implementation of the two different regression-based models \textcolor{black}{on the dataset of 87 waveforms}.

\subsubsection{MLR-based model}

The implementation of this model is quite straightforward since the model produced is \textcolor{black}{determined solely by} the regression coefficient matrix $\hat{B}$. Taking into account the fact that multivariate linear regression is \textcolor{black}{equivalent with} separate and independent multiple linear regressions (multi-input, single output),  \textcolor{black}{provided that responses are not correlated}, one can multiply $\hat{B}$ with any chosen column vector of four elements (that is, $\left[ 1, dR/dM, M, \kappa_2^\tau \right]$) to obtain an output vector that will represent the strain.

For the training, we consider the spectral amplitude in linear scale (in the various figures, we show the base-10 logarithm of the spectra). For the regressors, that is, the columns of the design matrix $X$, we performed a standardization of the values along each row of $X$.

\subsubsection{ANN-based model}

We used a 4-layer (three hidden plus output), feed-forward ANN, with sigmoid transfer functions (activations) in the hidden layers (comprising between 200 and 400 nodes each) and linear in its input and output ones. Additionally, an Adam optimizer with a batch size of 6 samples was used and training was completed in at most 100 epochs, due to early stopping. The optimizer itself introduces a small stochasticity in the final FF values between different trainings. We implemented the algorithm using the TensorFlow framework \cite{tensorflow2015-whitepaper}. The ANN architecture is summarized in Table \ref{tab:ann_arch}. To ensure the generalization of the proposed architecture, we also integrated layers adding Gaussian noise, as well as dropout layers (which also introduce small stochasticity).  The respective settings for these additional layers are also depicted in Table \ref{tab:ann_arch}. Similarly, Gaussian noise was added to the validation set. 

\begin{table}[bh!]
\caption{
Artificial neural network architecture, invoking the \textcolor{black}{summary function of the Tensorflow library \cite{tensorflow2015-whitepaper}}.
}
\label{tab:ann_arch}%
\begin{ruledtabular}
\resizebox{\linewidth}{!}{%
\begin{tabular}{ccccc}
Layer&Type&Shape&Activation&Params\\
\hline
\#1&GaussianNoise(0.1)&(None, 3)&-&0\\
\#2&Dense&(None, 200)&Linear&800\\
\#3&GaussianNoise(0.05)&(None, 200)&-&0\\
\#4&Dropout(0.15)&(None, 200)&-&0\\
\#5&Dense&(None, 400)&Sigmoid&80400\\
\#6&GaussianNoise(0.1)&(None, 400)&-&0\\
\#7&Dropout(0.15)&(None, 400)&-&0\\
\#8&Dense&(None, 400)&Sigmoid&160400\\
\#9&GaussianNoise(0.1)&(None, 400)&-&0\\
\#10&Dropout(0.05)&(None, 400)&-&0\\
\#11&Dense&(None, 370)&Linear&148370\\
\end{tabular}
}
\end{ruledtabular}
\end{table}

During training, we used an learning rate scheduler, which starts with a low LR (for warm-up) and increases the LR linearly in each step of the training process. The warm-up period lasted approximately 4 epochs. Finally, we scaled the input of the network using an established standardization technique and inverted the result of the output (unscale). \textcolor{black}{This standard scaling was also used for both the regressors and the regressands of the training set and of the validation set as well}. 

To train the network, we gave pairs of individual columns of the design matrix $X$ (regressors) in combination with the respective columns of the regressands matrix $Y$, which represented a specific spectrum. To be more precise, we feed the network with $X$'s column vectors omitting the first entry (with constant value of 1), since the bias is already included in the ANN  architecture. The hyperparameters were optimized heuristically. 

Furthermore, \textcolor{black}{to avoid any bottleneck in the architecture of the ANN, the consecutive layers have increasing number of nodes, except the last one having number of nodes equal to the frequency bins of the spectra}.

\begin{figure*}[ht!]
	\centering
	\begin{minipage}{0.5\textwidth}
		\centering
		\includegraphics[width=0.95\linewidth]{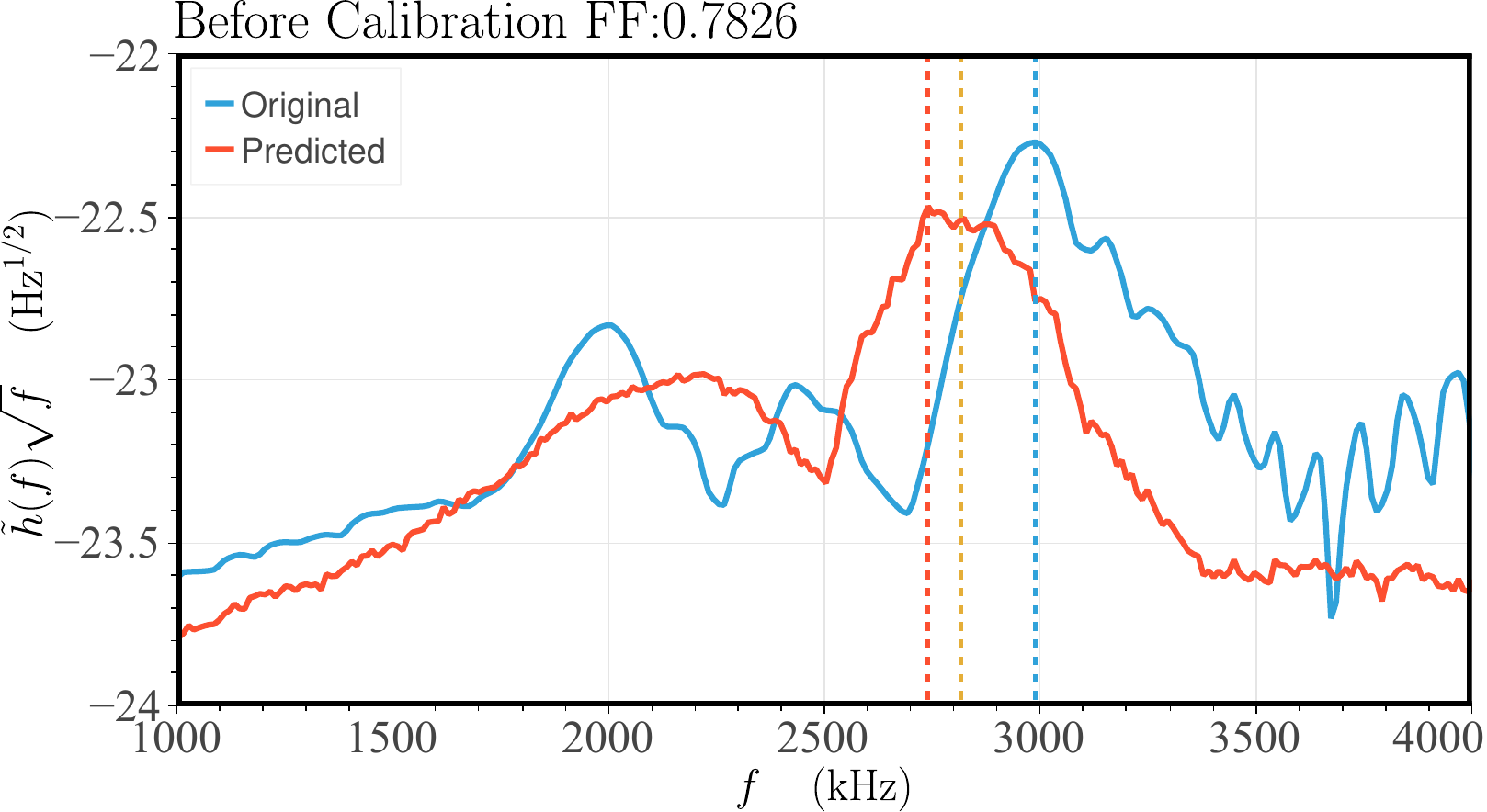}
	\end{minipage}%
	\begin{minipage}{0.5\textwidth}
		\centering
		\includegraphics[width=0.95\linewidth]{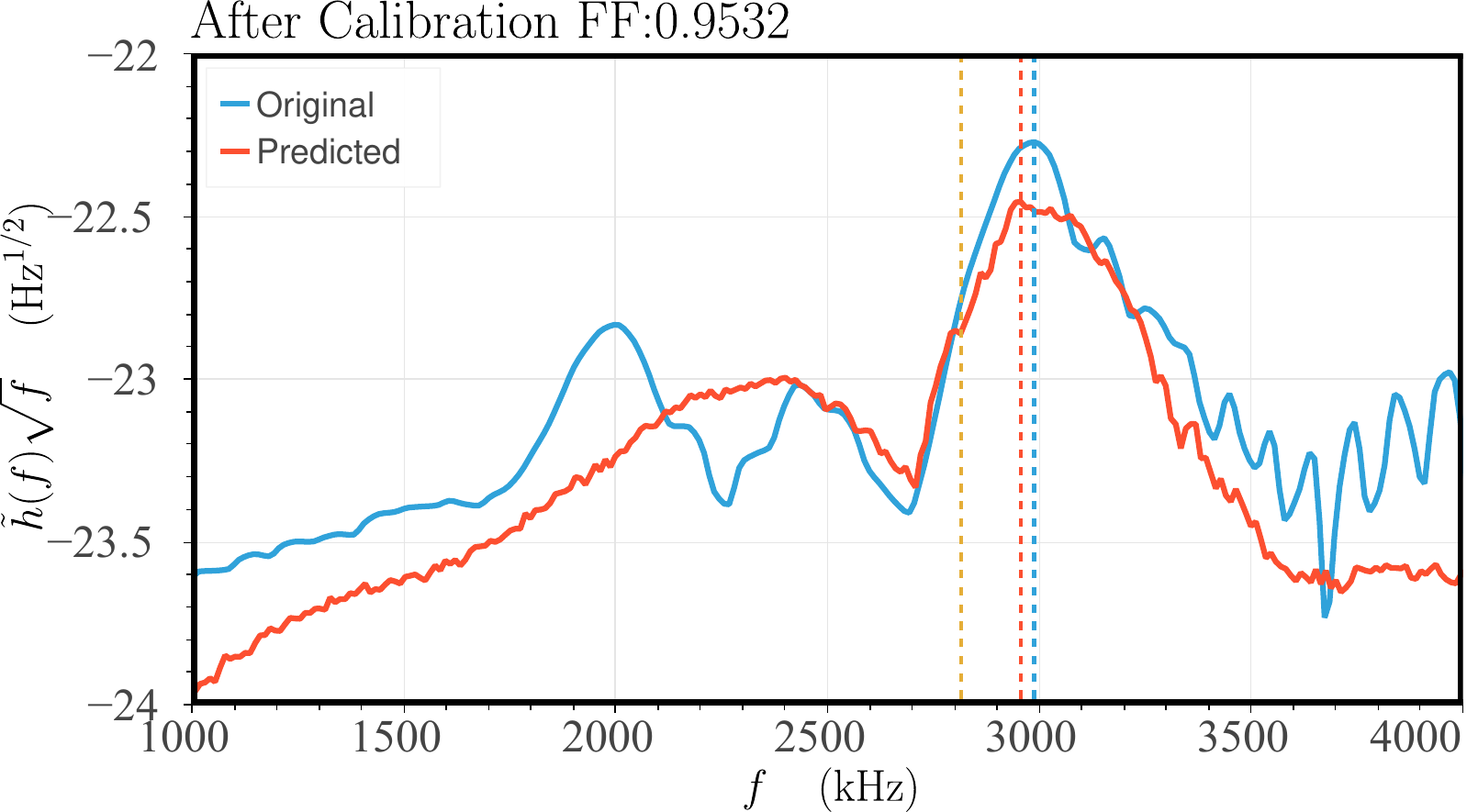}
	\end{minipage}
	\caption{Recalibration of the predicted postmerger spectrum for an 1.364$M_\odot$ equal-mass BNS merger with EOS LS220. The left panel shows the ANN-based prediction before calibration, where the main postmerger peak is misaligned with respect to the original spectrum. The right panel shows the calibrated spectrum, whis has a much better alignment, leading to a significantly higher fitting factor of 0.953.}
	\label{fig:LS220_recalibration}
\end{figure*}

In Figure \ref{fig:all_loss_curves} we depict \textcolor{black}{exemplary} learning curves for this ANN model to diagnose its behaviour. The validation loss curve follows the training loss curve and lies mostly below it, implying that the dataset is representative \cite{ANZANELLO2011573,goodfellow2016deep}, allowing for a good fit, from an ML perspective. We can also observe in these curves that the loss value does not exhibit a significant percentage decrease because it happens to be in the same range of initialised network weights (we used the default initialization in Tensorflow) \textcolor{black}{since we did not use any initialization function}. Lastly, the training procedure is stopped when the validation loss curve diverges for 7 epochs consecutively.

Before feeding the spectra into the ANN for the training procedure, we performed the same partial alignment (shift) procedure as for MLR. This \textcolor{black}{renders} structure in the training dataset (with respect to the regressands), not captured by the three variables we chose in the design matrix. 
This  may not have been necessary, if we had a larger training set and additional variables in the design matrix (e.g. including the maximum mass and the radius at the maximum mass). For example, in the time-domain model of \cite{PhysRevD.105.043020}, a larger number of parameters was used to achieve high fitting factors.

As a final remark, the ANN produces a predicted spectrum, by nonlinearly entangling information from all frequency bins, which is not the case for the MLR model, where at each frequency bin, a linear regression, independent of other frequency bins, is performed. 

\subsection{Recalibration of predicted spectra}
\label{sec:recalibration}

Since we are using an empirical relation for the partial alignment of the spectra, the inverse procedure induces an error in the predicted spectrum that depends on the accuracy of the empirical relation. To test the impact of this error, we performed a "recalibration", as in  \cite{breschi2022numerical}. In this way, one can reverse, to some extent, the horizontal or vertical mis-alignement of a predicted spectrum with respect to the original one. This is useful, as it shows the real potential of our method for predicting postmerger spectra in the future, when more training samples will be available and, in addition, the empirical relation will become more accurate, due to tightening EOS constraints derived from astrophysical observations (see, e.g. \cite{2024_nature}).

In \cite{breschi2022numerical}, several parameters were used for recalibration. Here, we only use the known peak frequency of the postmerger spectra in the training set. We therefore define as $\Delta f_\mathrm{peak}$ the difference between the peak frequency of the original spectrum and the peak frequency of the predicted spectrum after calibration. Then, we define the likelihood function
\begin{equation}
    \mathcal{L}(\Delta f_\mathrm{peak}) = { \sum_{f_\mathrm{min}}^{f_\mathrm{max}}}{\left[( h_{\rm o}(f) - h_{\rm p}(f - \Delta f_\mathrm{peak}) \right]^{2}},
    \label{eq:likelihood}
\end{equation}
where $h_{\rm o}$ is the original strain spectrum, $h_{\rm p}$ the predicted strain spectrum, while $f_\mathrm{min}$ and $f_\mathrm{max}$ define the frequency range for recalibration.
This likelihood is basically a mean-squared-error (MSE) function that is minimized close to the value of $\Delta f_\mathrm{peak}$,
through the Maximum Likelihood Estimation (MLE) technique. This is equivalent to Bayesian inference using uninformed priors. In practice, we use 60 frequency values in the range 1-4 kHz. An indicative example of recalibration is shown in Figure \ref{fig:LS220_recalibration}. 

In Appendix \ref{AppendixA} we display collective plots of predicted vs. original spectra, produced using a $k-$fold CV with $k=4$ (each plot shows a hold-out case, not included in the corresponding training sample)\footnote{ The choice of $k=4$ is common in such investigations.}, see Figs. \ref{fig:Takami_neural_cl} to \ref{fig:Soultanis_neural_cl}. In all figures, the calibrated cases are shown. The predicted spectra are, in most cases, close to the predicted ones, even in the low- and high-frequency tails in this part of the postmerger spectrum. There is good alignment of the main frequency peaks and in most cases the shape of one or more secondary peaks is also predicted with satisfactory accuracy. In some cases, the main peak in the original spectra is quite narrow, which is due to a long-lived $f_2$ oscillation. This feature cannot yet be captured by our 3-parameter model and the predicted spectra show a somewhat wider peak, with smaller maximum amplitude. This could be remedied in future improved models, by including more parameters.

\subsection{Distribution of fitting factors}

\begin{figure*}
\centering
\begin{minipage}{0.5\textwidth}
  \centering
  \includegraphics[width=0.95\linewidth]{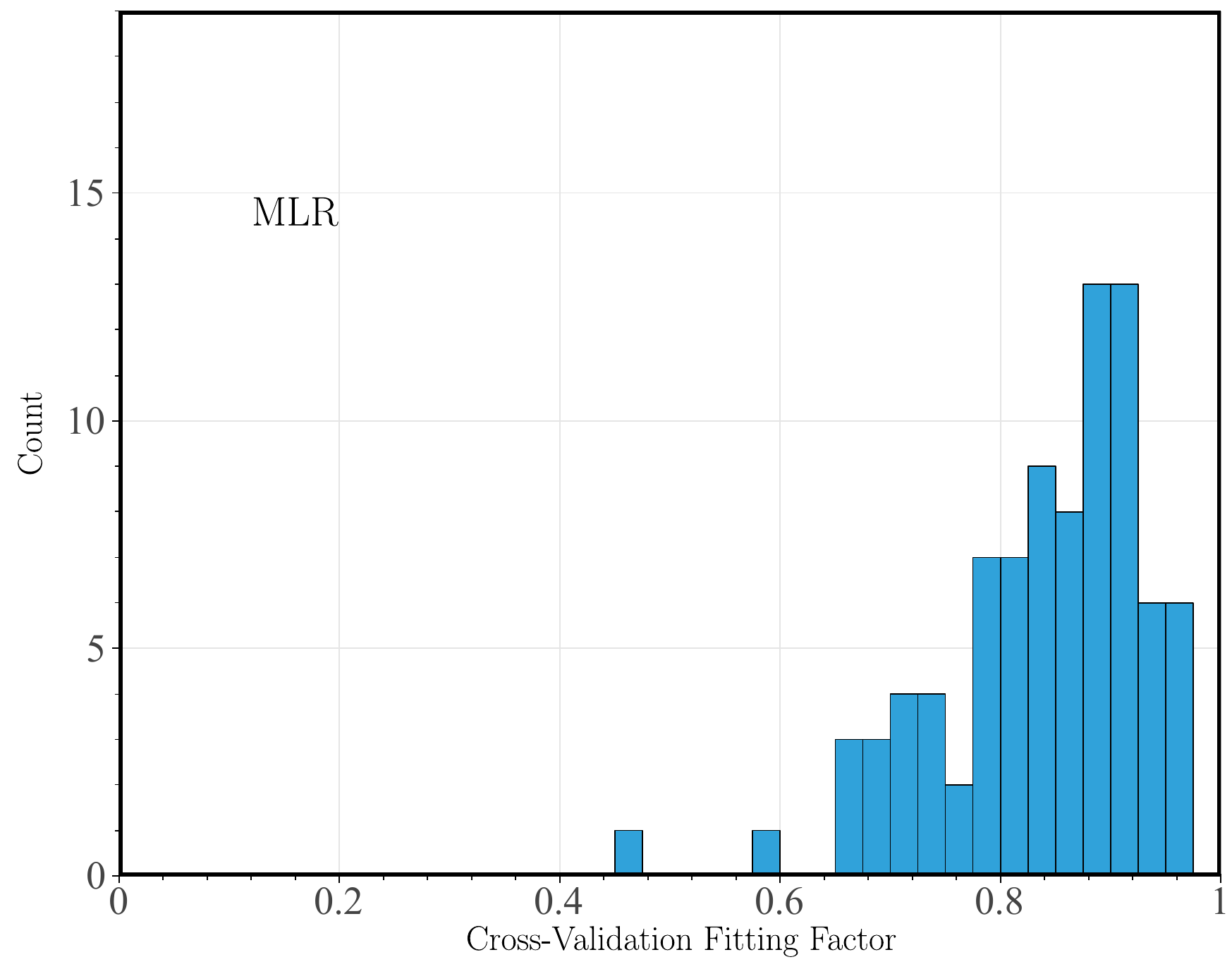}
\end{minipage}%
\begin{minipage}{0.5\textwidth}
  \centering
  \includegraphics[width=0.95\linewidth]{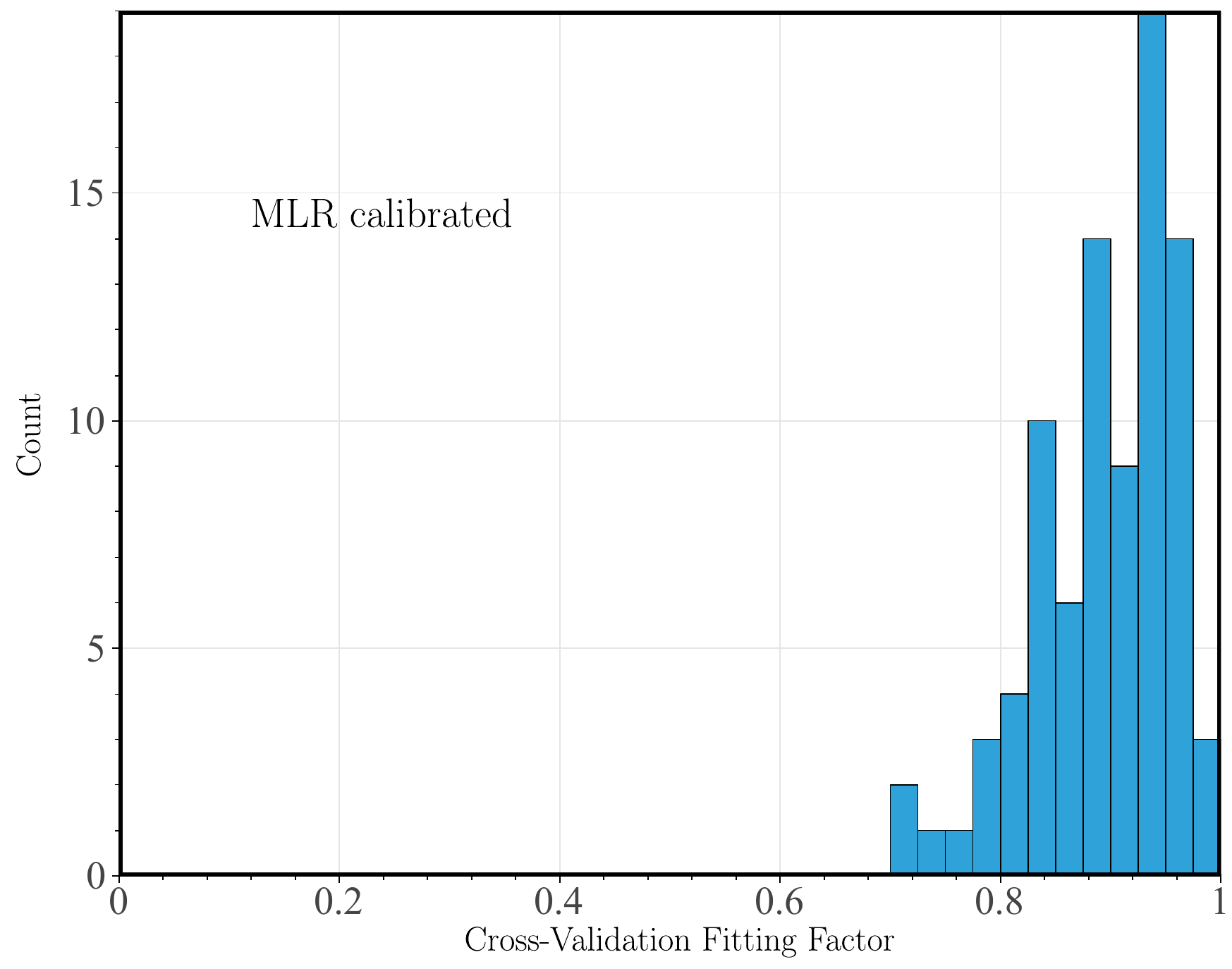}
\end{minipage}
\begin{minipage}{0.5\textwidth}
  \centering
  \includegraphics[width=0.95\linewidth]{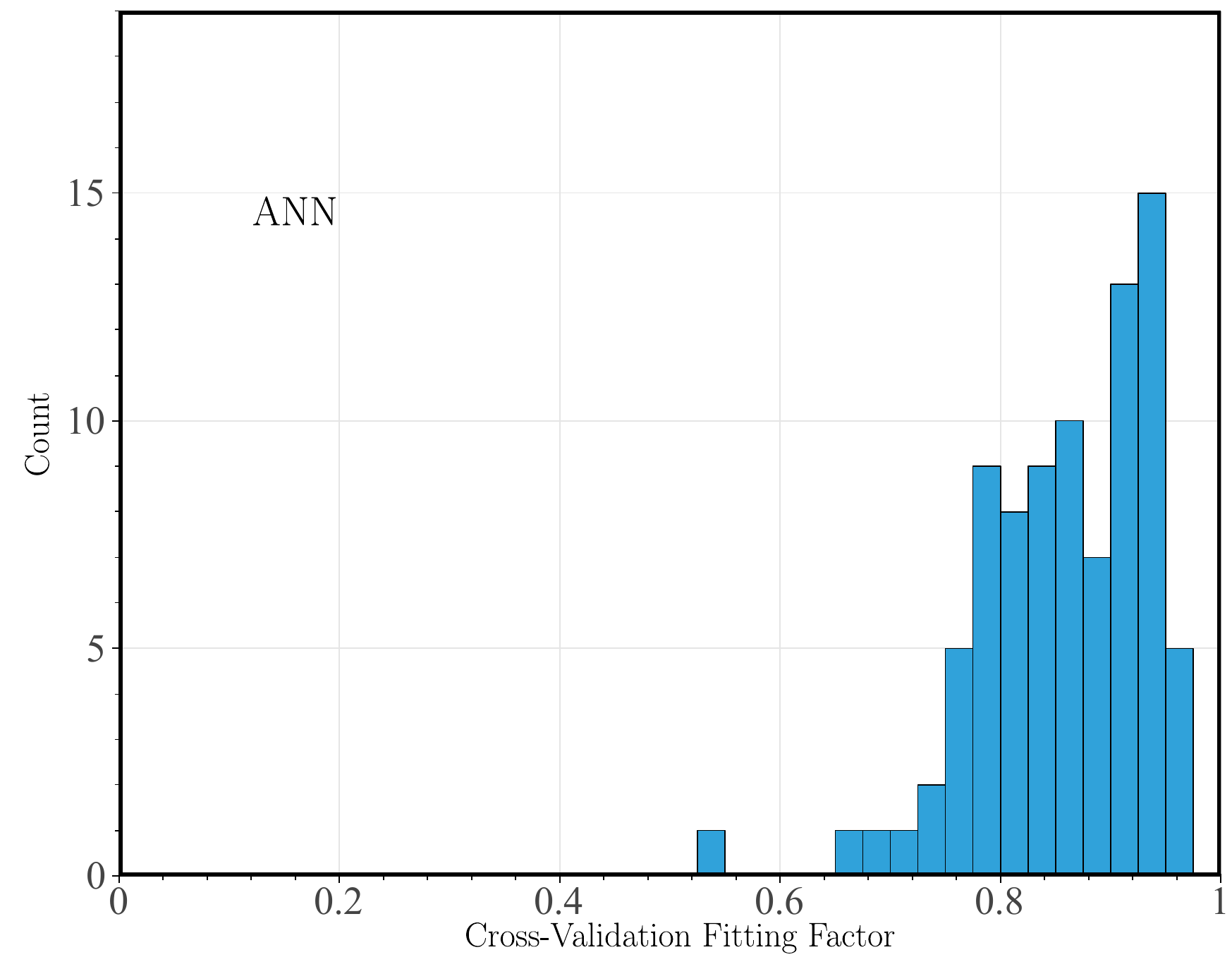}
\end{minipage}%
\begin{minipage}{0.5\textwidth}
  \centering
  \includegraphics[width=0.95\linewidth]{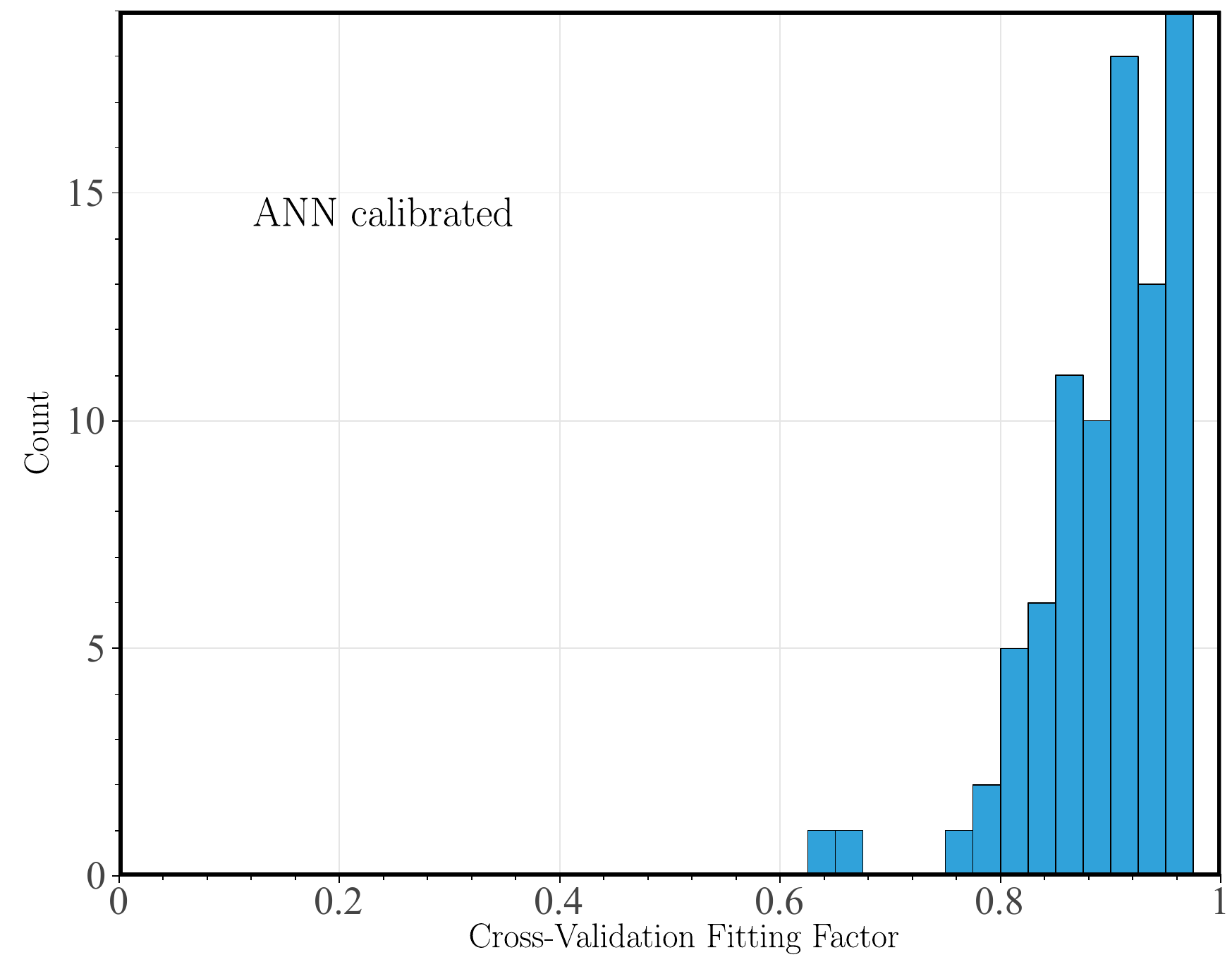}
\end{minipage}%
\caption{Histograms of fitting factors for predicted postmerger spectra when using the  multivariate linear regression (MLR)  (top row) and the ANN model (bottom row), for $k=4$ cross validation. The left column corresponds to the cases without recalibration, while in the right column recalibration was applied (see the text for details). The distribution of the fitting factors obtained with the ANN models is \textcolor{black}{more concentrated towards one} than for the MLR model.}
\label{fig:histograms}
\end{figure*}

In the left column of Figure \ref{fig:histograms}, the calculated histograms of the fitting factors FF are depicted in the uncalibrated case for the MLR and ANN models, respectively. We set the number of bins to 40 in each case, and the number $k$ of CV folds equal to 4.

In the uncalibrated case, we observe that the fitting factor distributions \textcolor{black}{of} the two models have a similar shape, but the histogram is more concentrated to higher values for the ANN model. Specifically, for the MLR model there are 18 fitting factor values below 0.8, whereas for the ANN model, this reduces to only 11. Moreover, the mode of the ANN histogram is closer to 1, and there are fewer outliers. 

On the other hand, after recalibration both histograms greatly improve, \textcolor{black}{as shown in} the right column of Figure \ref{fig:histograms}. The mode increases, the histogram becomes narrower with values closer to one, and there are even fewer outliers, compared to the uncalibrated case. Comparing the MLR calibrated histogram of fitting factors with the ANN calibrated one, we observe that the latter \textcolor{black}{is more concentrated closer to one}, except for two outliers.

\begin{figure*}
\centering
\begin{minipage}{0.5\textwidth}
  \centering
  \includegraphics[width=\linewidth]{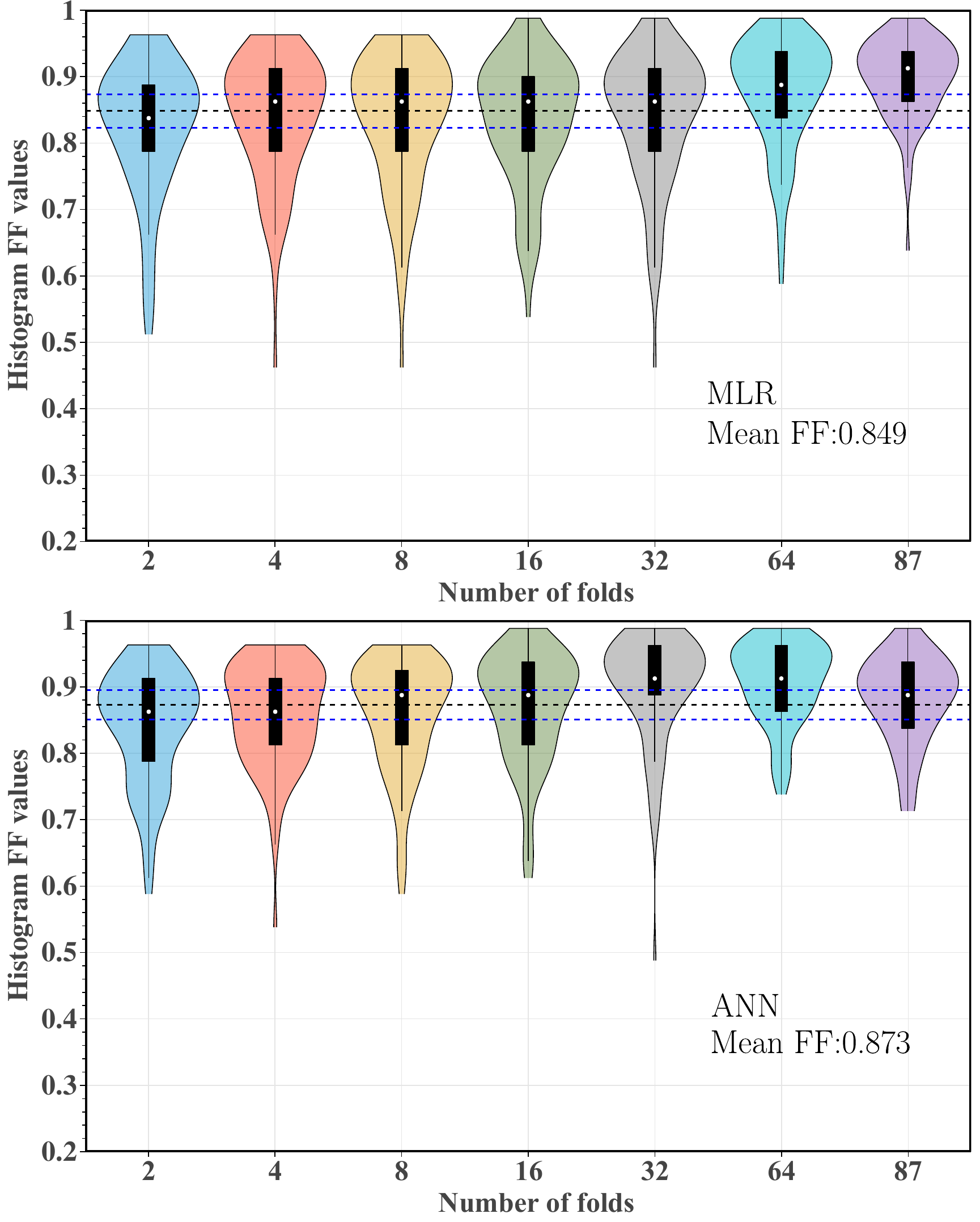}
\end{minipage}%
\begin{minipage}{0.5\textwidth}
  \centering
  \includegraphics[width=\linewidth]{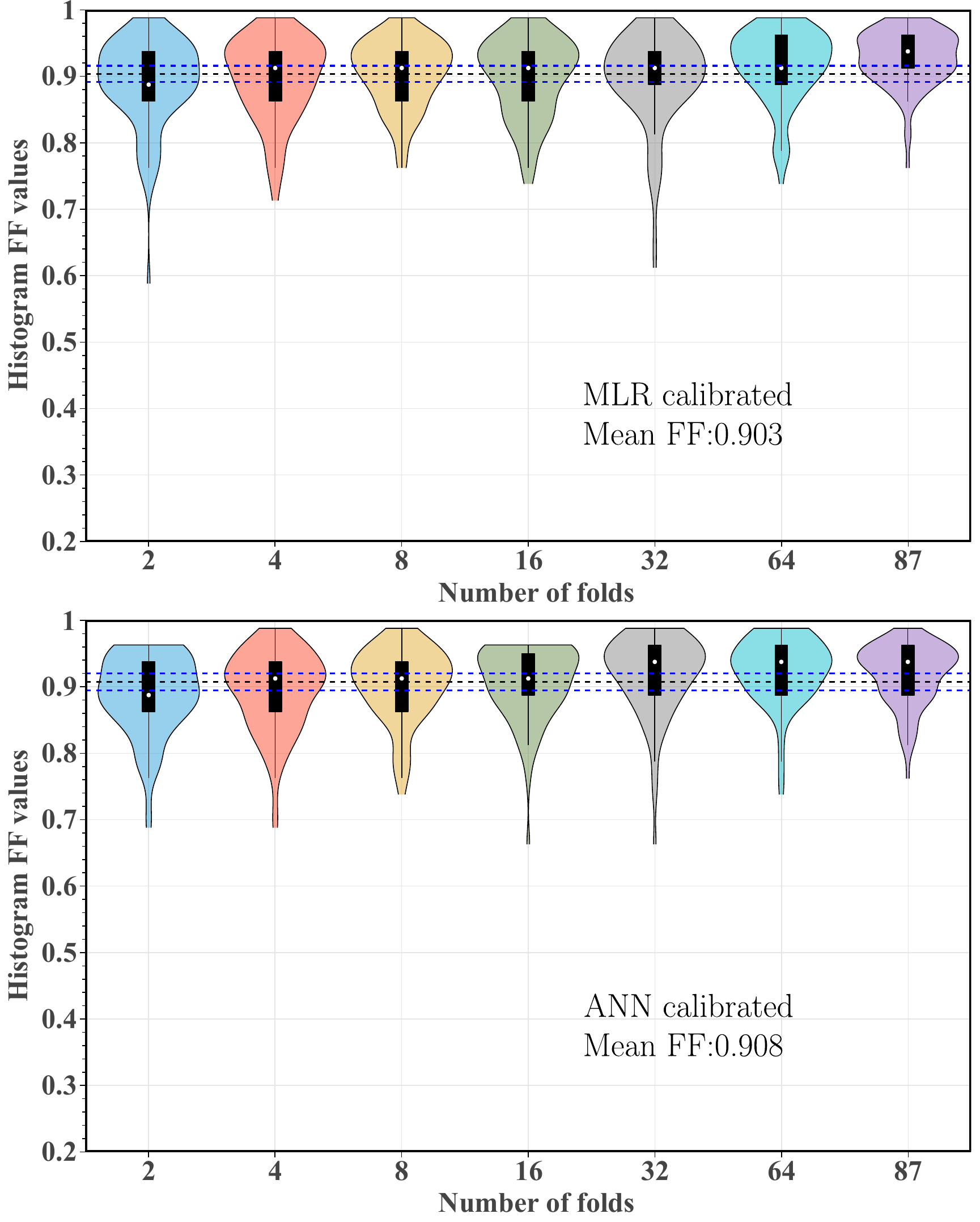}
\end{minipage}
\caption{Violin plot representations of the distribution of fitting factors for the MLR model (top row) and the ANN one (bottom row) for different number of folds $k$ in $k-$fold cross-validation. 
The left column corresponds to the cases without recalibration, while in the right column, recalibration was applied (see text for details). The mean fitting factor among all the number of folds is shown as a dashed black horizontal line and the two dashed blue horizontal lines represent one standard deviation. The highest mean fitting factor (0.908) is obtained with the ANN calibrated model.}
\label{fig:violin_plots_per_N}
\end{figure*}

In addition to the comparison of the fitting factor histograms, we also investigate the behavior of the models with respect to changing the value of $k$ in the CV k-fold. In Fig. \ref{fig:violin_plots_per_N}  the histograms of the fitting factors are displayed as separate violin plots.   We observe that irrespective of individual cases of $k$, the overall behavior of the ANN models is slightly better than that of the MLR, since the mean FF value across all $k$'s is higher and less dispersed. Regarding the dispersion of the FF values in the MLR case, we observe that the inter-quartile range (IQR) decreases slightly as the number of folds $k$ increases and becomes minimum when using leave-one-out cross validation, leading to concentration of values in higher histogram bins (bias-variance trade-off). The ANN model is more stable in terms of dispersion of FF values and exhibits similar behaviour when varying $k$. The MLR model is significantly more susceptible to outliers than the ANN model. We note that to make the comparisons, we first performed a stable shuffling in the training dataset for both the regressors and the regressands.

\section{CONCLUSIONS} \label{section_iv}

In this work, we presented two robust methods for predicting the GW postmerger spectrum for binary neutron star mergers. 
One method uses multivariate least-squares regression \textcolor{black}{(MLR)}, whereas the second method \textcolor{black}{uses} artificial neural networks \textcolor{black}{(ANN)}. 
 
Concerning the multivariate least-squares regression, our work improves on the previous work by Easter et al. \cite{PhysRevD.100.043005} in \textcolor{black}{the following points}: a) in the design matrix, we use the inverse gradient $dR/dM$ of specific points in the $M(R)$ plots instead of the compactness $C=M/R$. b) We use a refined empirical relation for the postmerger peak frequency in terms of other characteristics of neutron stars, presented in Vretinaris et al. \cite{PhysRevD.101.084039_Vretinaris}, instead of a different relation used in \cite{PhysRevD.100.043005}. c) We use an extended dataset of 87 different spectra, compared to 35 in \cite{PhysRevD.100.043005}.
These changes lead to higher fitting factors of the predicted spectra. Our approach is robust, taking into consideration that we used an unbalanced design matrix.

In addition, we demonstrate that a trained ANN can predict the postmerger spectra with higher mean fitting factors compared to the MLR model. Specifically, we 
used a 4-layer feed-forward ANN (with the Adam optimizer), comprising between 200 and 400 nodes in each layer and with additional Gaussian noise layers and dropout ones (which introduce a small stochasticity).  

We evaluated the accuracy of the two methods in predicting postmerger spectra using the standard fitting factor and performed several $k$-fold cross-validations. For both the $k=4$ case (which corresponds approximately to a 75\%-25\% train/test split), as well as the $k=87$ case (which corresponds to the leave-one-out method) we find that the histogram of fitting factors is comparable between the two methods, when calibration is performed. However, we note a slightly higher mean fitting factor (across all $k$-folds) when using the ANN-based approach.

The distribution of fitting factors when using the recalibrated spectra is significantly improved with respect to the uncalibrated predictions. Essentially, the recalibrated results remove, to some degree, the uncertainties introduced by the use of the empirical relation in aligning the spectra during training. We note that postmerger spectra are realistically expected to be observed with 3rd-generation detectors, which will also significantly constrain the EOS using information from the inspiral phase; see, e.g., the anticipated constraints in \cite{iacovelli2023nuclear}, when assuming 500 BNS observations with 3rd-generation detectors. In parallel, the available collections of postmerger BNS spectra produced by numerical-relativity simulations is expected to be significantly enlarged in the time it will take for 3rd-generation detectors to become operational. These two improvements will allow us to significantly reduce the uncertainty of the empirical relation, and thus our calibrated results point towards the anticipated accuracy of our method at the time when actual observations will \textcolor{black}{be obtained}.

Possible future improvements include the extension of the design matrix by adding, e.g., quadratic terms or additional physical characteristics. Furthermore, the design matrix could take into account specific EOS information, leading to categorical regression.

The methods presented here can be used to create template banks of postmerger spectra, which will be useful for detecting this phase after a BNS event. On the other hand, the methods could be inverted, allowing the estimation of the parameters included in the design matrix, given a potential observation (see also \cite{PhysRevD.100.043005}). We are planning to investigate this application in forthcoming work.

In this initial investigation, we have examined a basic configuration for the BNS simulations, focusing solely on hydrodynamics while disregarding the influence of magnetic fields, neutrino transport and other forms of dissipative effects or effective viscosities 
\cite{Kiuchi:2014hja,Giacomazzo:2014qba,Kiuchi:2015sga,Radice:2017zta,Kiuchi:2017zzg,Ciolfi:2019fie,PhysRevD.105.104028,2022PhRvD.105j3020P,2023PhRvD.107d4037W,2023PhRvD.107j4059E,2023LRCA....9....1F,2024arXiv240102493M}.
It will be important to conduct comprehensive analyses that incorporate all relevant physical factors to have a faithful representation of the postmerger spectra. In addition, more work is needed to produce a large number of training spectra with unequal masses. Finally, future studies will need to include the potential impact of departures from general relativity on the postmerger spectra produced during BNS mergers.

\begin{acknowledgments}
We are truly grateful to Paul Easter for providing us with notes on his previous publication on this subject, as well as to Nikolaos Karnesis and Paraskevi Nousi for their insight into the spectral recalibration procedure and the ANN architecture, respectively. We would additionally like to thank the members of the Extreme Matter Working Group of Virgo for the opportunity they gave us to present our work, and \textcolor{black}{Tim Dietrich} and Andreas Bauswein and Bruno Giacomazzo for their comments on the manuscript. DP would finally like to thank the Holoviews community for providing help on how to export the various plots in SVG format, through their discourse platform.

This research work was supported by the Hellenic Foundation for Research and Innovation (HFRI) under the 3rd Call for HFRI PhD Fellowships (Fellowship Number: 6836). We acknowledge support by Virgo, which is funded, through the European Gravitational Observatory (EGO), by the French Centre National de Recherche Scientifique (CNRS), the Italian Istituto Nazionale di Fisica Nucleare (INFN) and the Dutch Nikhef, with contributions by institutions from Belgium, Germany, Greece, Hungary, Ireland, Japan, Monaco, Poland, Portugal, Spain.
\end{acknowledgments}

\appendix

\section{Collective plots of models predictions} \label{AppendixA}

\begin{figure*}[ht!]
    \centering
    \includegraphics[height=0.95\textheight,width=\textwidth]{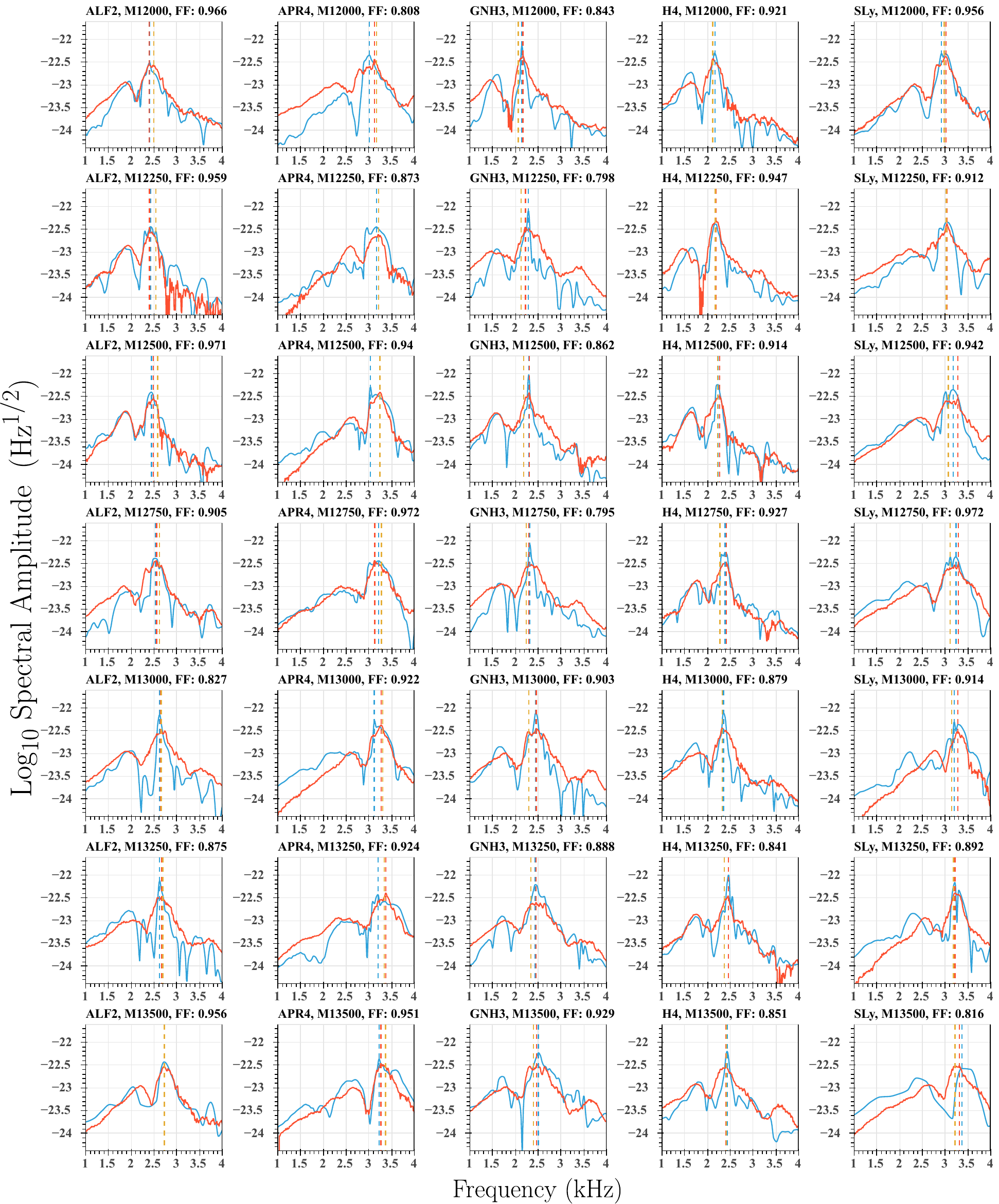}
    \caption{Original  gravitational-wave spectra of BNS mergers (blue curves) depicted along with their ANN-based model predicted ones (red curves) for the Rezzolla et al. dataset of numerical simulations \cite{PhysRevD.100.043005}, except for the H4 and SLy cases with component mass of 1.35 $M_\odot$, which were chosen from the CoRe database \cite{Gonzalez_2022mgo}. The blue and red dashed vertical lines correspond to the peak value of the original and the predicted spectra, correspondingly. The orange dashed vertical line represents the prediction of the empirical relation Eq. (\ref{eq:empiricalVret}). }
    \label{fig:Takami_neural_cl}
\end{figure*}

\begin{figure*}[ht!]
    \centering
     \includegraphics[height=\textheight]{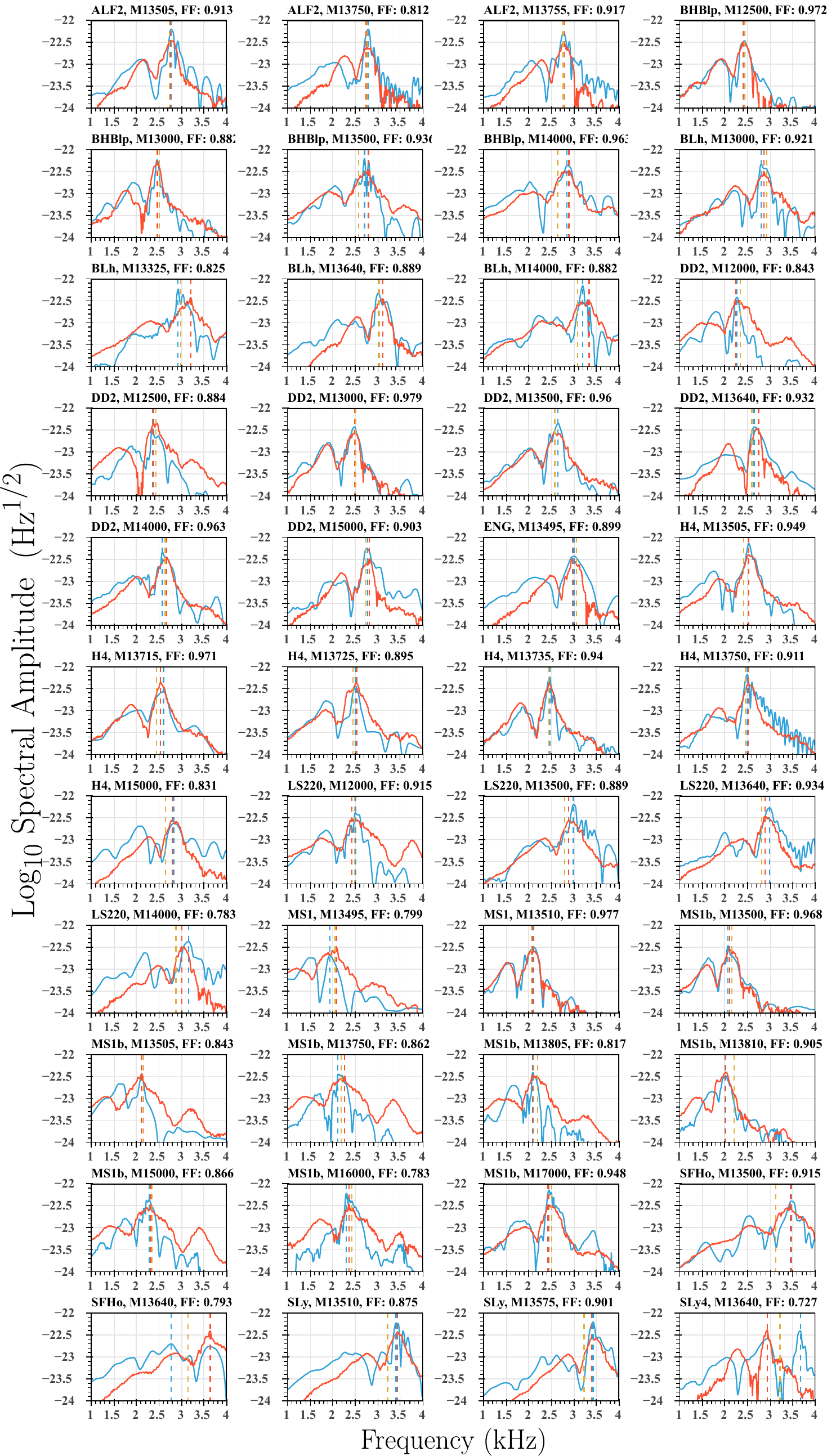}
    \caption{Same as Fig. \ref{fig:Takami_neural_cl}, but for models included in the CoRe \cite{Gonzalez_2022mgo} dataset.}
    \label{fig:Core_neural_cl}
\end{figure*}

\begin{figure*}[ht!]
    \centering
    \includegraphics[width=\textwidth,keepaspectratio]{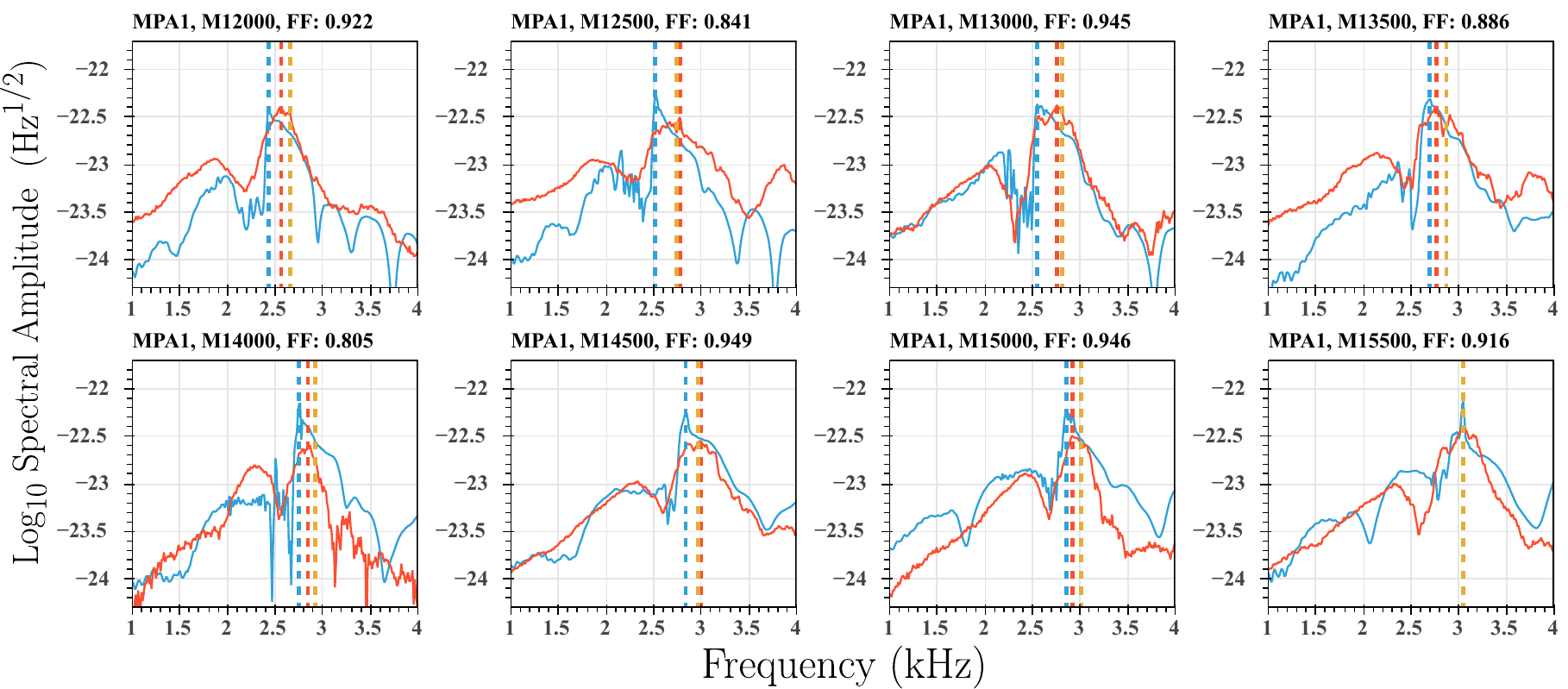}
    \caption{Same as Fig. \ref{fig:Takami_neural_cl}, but for models from Soultanis et al. \cite{PhysRevD.105.043020}.}
    \label{fig:Soultanis_neural_cl}
\end{figure*}

In Figures \ref{fig:Takami_neural_cl} to \ref{fig:Soultanis_neural_cl} individual predictions of the ANN-based model are depicted for each EOS and mass using 4-fold CV. 

In particular, in Figure \ref{fig:Takami_neural_cl} we present the predictions for the Rezzolla et al. data set \cite{PhysRevD.100.043005} of numerical simulations, except for the last row, that is, H4 and SLy EOS for the component mass 1.35 $M_\odot$, where we used simulations from the CoRe dataset \cite{Gonzalez_2022mgo}.  Similarly, Figures \ref{fig:Core_neural_cl} and \ref{fig:Soultanis_neural_cl} collectively represent predictions for spectra that were taken from the CoRe dataset \cite{Gonzalez_2022mgo} and Soultanis et al. \cite{PhysRevD.105.043020}, respectively. In these plots the blue and red dashed vertical lines correspond to the peak value of the original and the predicted spectra, respectively, whereas the orange dashed vertical line represents the prediction of the empirical relation Eq. (\ref{eq:empiricalVret}).

\begin{figure*}[ht!]
    \centering
    \includegraphics[height=0.95\textheight,width=\textwidth]{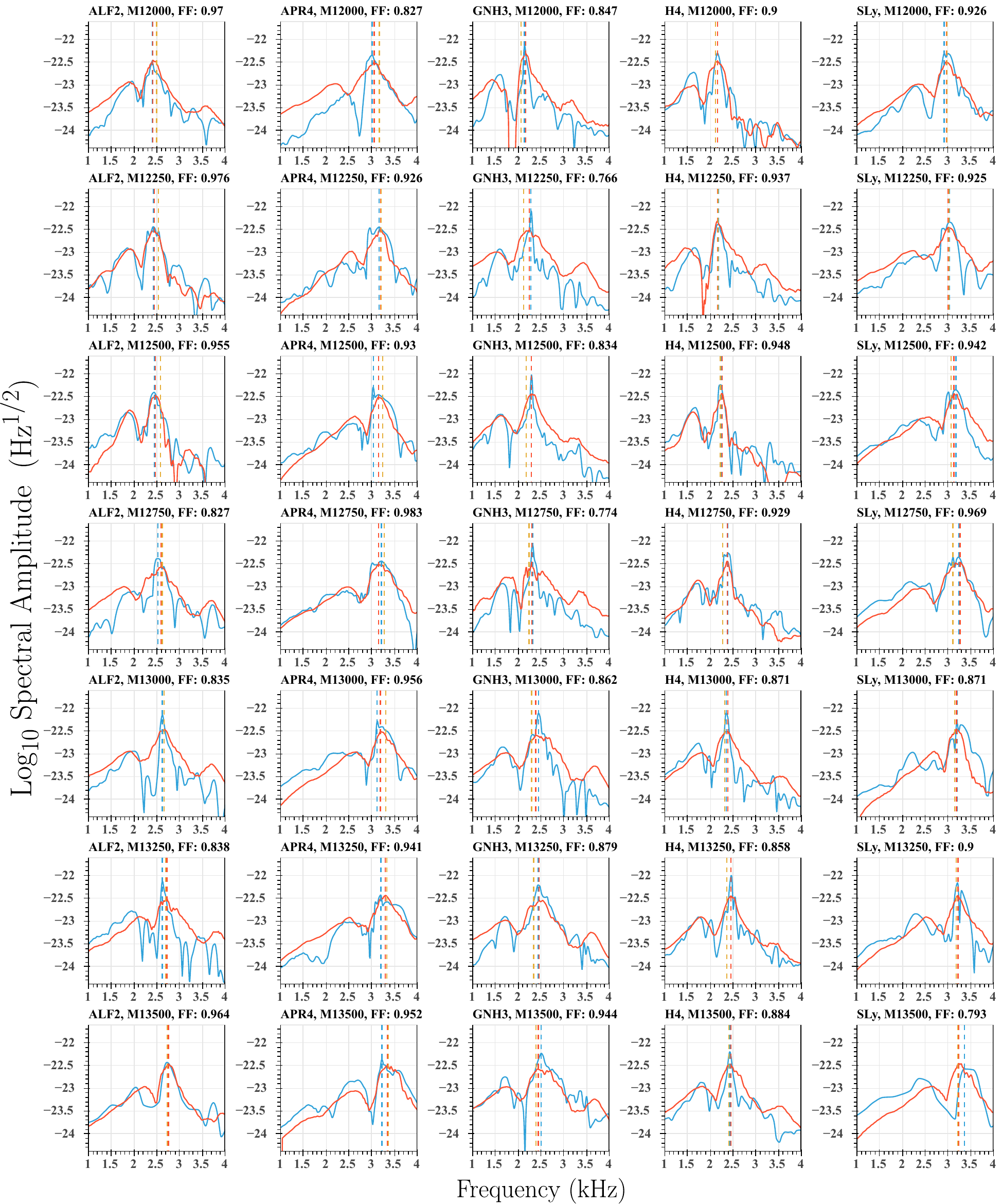}
    \caption{Same as Fig. \ref{fig:Takami_neural_cl}, but using the MLR-based model.}
    \label{fig:Takami_linreg_cl}
\end{figure*}

\begin{figure*}[ht!]
    \centering
\includegraphics[height=\textheight]{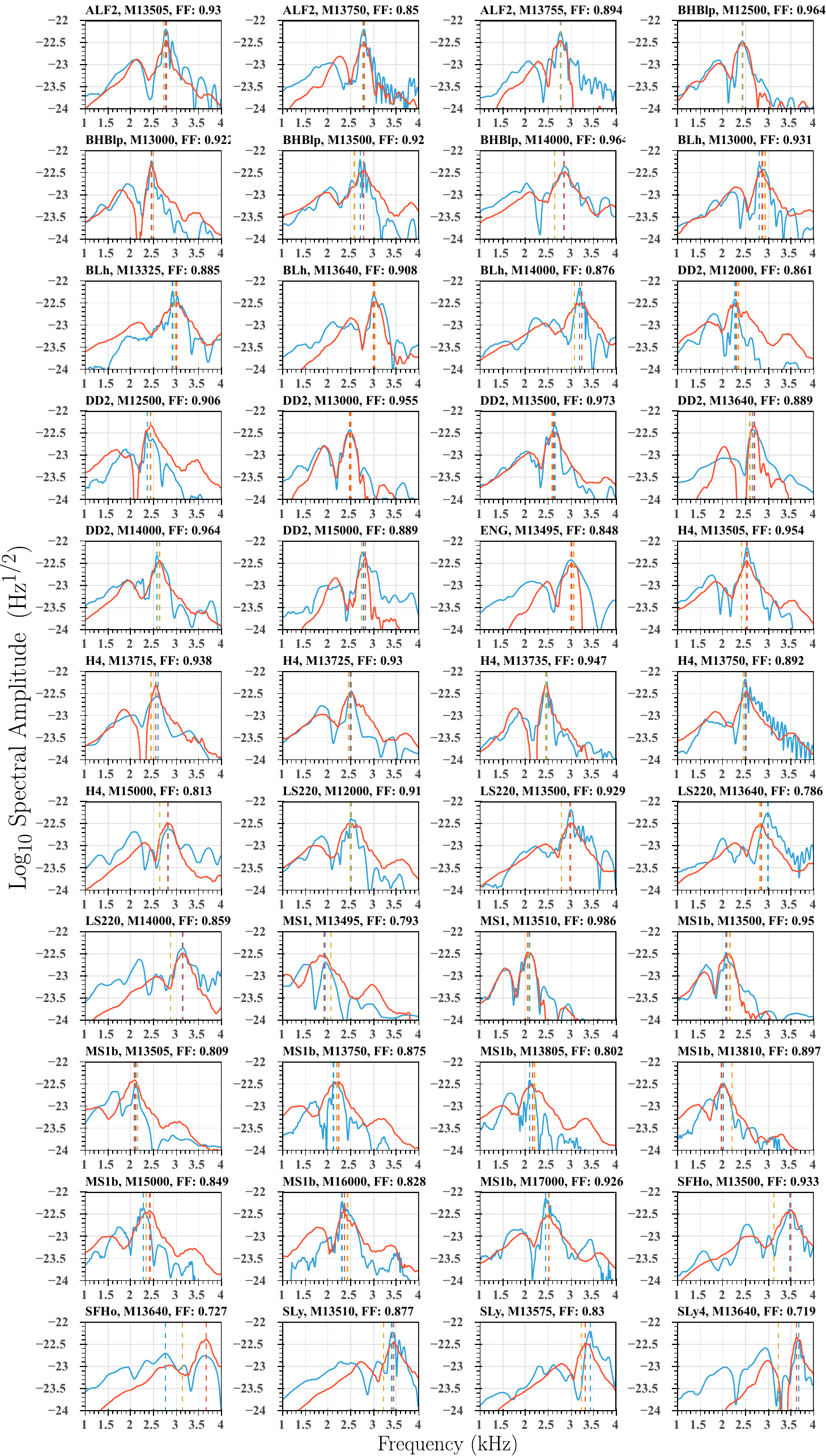}
    \caption{Same as Fig. \ref{fig:Core_neural_cl}, but using the MLR-based model.}
    \label{fig:Core_linreg_cl}
\end{figure*}

\begin{figure*}[ht!]
    \centering
\includegraphics[width=\textwidth,keepaspectratio]{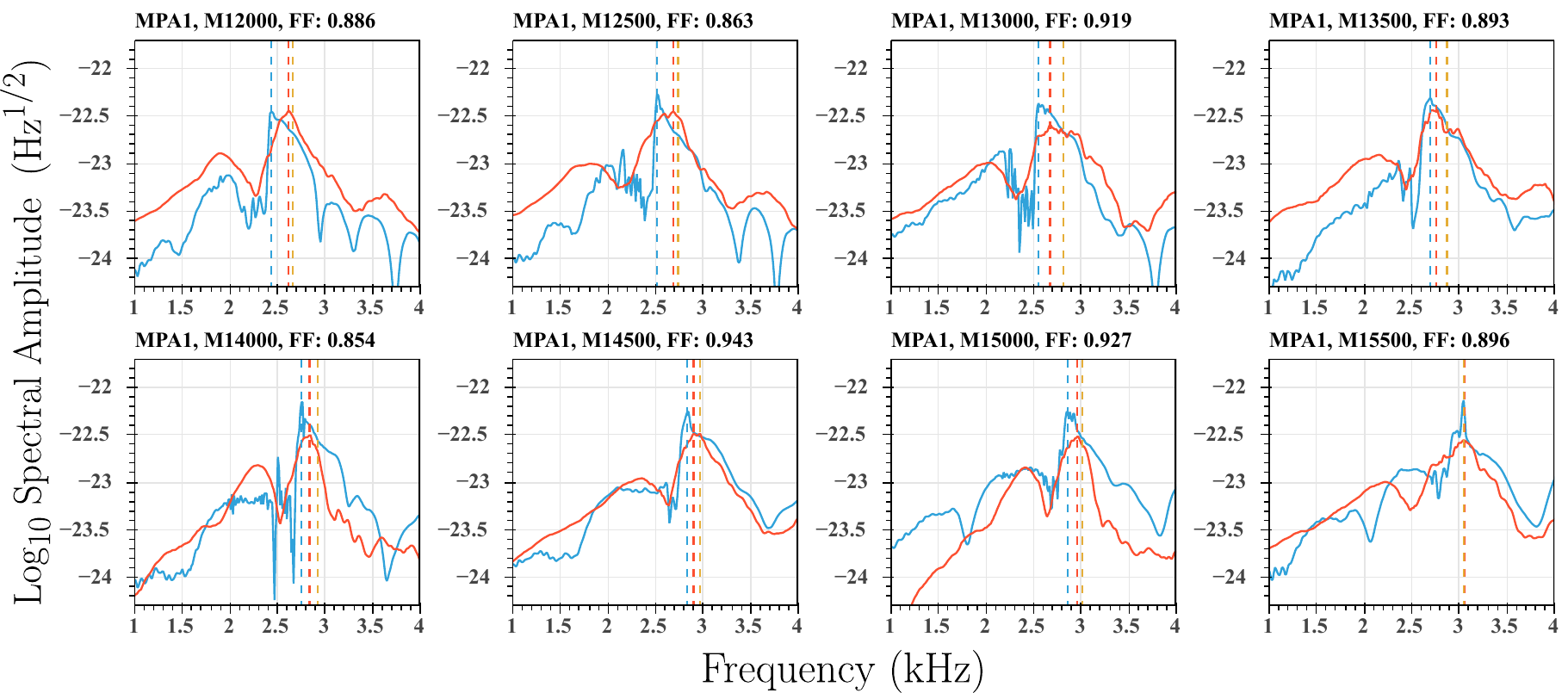}
    \caption{Same as Fig. \ref{fig:Soultanis_neural_cl}, but using the MLR-based model.}
    \label{fig:Soultanis_linreg_cl}
\end{figure*}

Finally, Figures \ref{fig:Takami_linreg_cl} to \ref{fig:Soultanis_linreg_cl} show the corresponding spectra when using the MLR-based approach. 

\pagebreak

\pagebreak


\pagebreak

\pagebreak
\pagebreak

\providecommand{\noopsort}[1]{}\providecommand{\singleletter}[1]{#1}%

\pagebreak

\end{document}